\documentclass{aa}  
\usepackage[dvipsnames,svgnames]{xcolor}
\usepackage{graphicx}
\usepackage{txfonts}
\usepackage{tabularx}
\usepackage[switch]{lineno}
\usepackage{amsmath,siunitx}
\usepackage{hyperref}
\bibstyle{aa}
\usepackage{natbib}
\usepackage{orcidlink}
\usepackage{booktabs}
\usepackage{makecell}
\hypersetup{
    colorlinks=true,
    citecolor=blue,
    urlcolor=blue,
    linkcolor=purple,
}
\titlerunning{An Eta Carinae Analog in the Sunburst Arc}
\authorrunning{S. Choe et al.}

\newcommand{\lya}{Ly$\alpha$}
\newcommand{\lyb}{Ly$\beta$}
\newcommand{\lam}{$\lambda$}
\newcommand{\kms}{$\rm{km\,s^{-1}}$}

\begin{document} 

   \title{The Sunburst Arc with JWST: II. Observations of an Eta Carinae Analog at $z=2.37$}


   \author{S. Choe\orcidlink{0000-0003-1343-197X}\inst{1}\thanks{Corresponding author; \email{suhyeon.choe@astro.su.se}}
          \and
          T. Emil Rivera-Thorsen\inst{1}\orcidlink{0000-0002-9204-3256}
          \and
          H. Dahle\inst{2}\orcidlink{0000-0003-2200-5606}
          \and 
          K. Sharon \inst{3}\orcidlink{0000-0002-7559-0864}
          \and 
          M. Riley Owens \inst{4}\orcidlink{0000-0002-2862-307X}
          \and
          J. R. Rigby \inst{5}\orcidlink{0000-0002-7627-6551}
          \and 
          M. B. Bayliss \inst{4}\orcidlink{0000-0003-1074-4807}
          \and 
          M. J. Hayes\inst{1}\orcidlink{0000-0001-8587-218X}
          \and 
          T. Hutchison \inst{5}\orcidlink{0000-0001-6251-4988}
          \and 
          B. Welch\inst{5,6,7}\orcidlink{0000-0003-1815-0114}
          \and
          J. Chisholm\inst{8}\orcidlink{0000-0002-0302-2577}
          \and 
          M. D. Gladders \inst{9,10}\orcidlink{0000-0003-1370-5010}
          \and 
          G. Khullar \inst{11}\orcidlink{0000-0002-3475-7648}
          K. Kim
          \inst{12}\orcidlink{0000-0001-6505-0293}
          }

   \institute{The Oskar Klein Centre, Department of Astronomy, Stockholm University, AlbaNova 10691, Stockholm, Sweden
        \and
        Institute of Theoretical Astrophysics, University of Oslo, P.O. Box 1029, Blindern, NO-0315 Oslo, Norway
        \and
        Department of Astronomy, University of Michigan, 1085 S. University Ave, Ann Arbor, MI 48109, USA
        \and
        Department of Physics, University of Cincinnati, Cincinnati, OH 45221, USA
        \and
        Astrophysics Science Division, Code 660, NASA Goddard Space Flight Center, 8800 Greenbelt Rd., Greenbelt, MD 20771, USA
        \and
        Department of Astronomy, University of Maryland, College Park, MD 20742, USA
        \and
        Center for Research and Exploration in Space Science and Technology, NASA/GSFC, Greenbelt, MD 20771
        \and
        Department of Astronomy, University of Texas at Austin, 2515 Speedway, Austin, Texas 78712, USA
        \and
        Department of Astronomy and Astrophysics, University of Chicago, 5640 South Ellis Avenue, Chicago, IL 60637, USA
        \and
        Kavli Institute for Cosmological Physics, University of Chicago, Chicago, IL 60637, USA
        \and
        Department of Physics and Astronomy and PITT PACC, University of Pittsburgh, Pittsburgh, PA 15260, USA
        \and
        IPAC, California Institute of Technology, 1200 E. California Blvd., Pasadena CA, 91125, USA
             }

   \date{Received April 2024}

 
  \abstract
  %
  %
  {``Godzilla'' is a peculiar object within the gravitationally lensed Sunburst Arc at $z=2.37$. Despite being very bright, it appears in only one of the twelve lensed images of the source galaxy, and shows exotic spectroscopic properties not found in any other clumps.}
   {We use JWST's unique combination of spatial resolution and spectroscopic sensitivity to propose a unified, coherent explanation of the physical nature of Godzilla.}
   %
   %
   {We measure fluxes and kinematic properties of rest-optical emission lines in Godzilla and surrounding regions. Using standard line ratio-based diagnostic methods in combination with NIRCam imaging and ground based rest-UV spectra, we characterize Godzilla and its surroundings.}
   {
   Among around 60 detected lines, we find a cascade of strong \ion{O}{i} lines pumped by intense \lyb\ emission, as well as \lya-pumped rest-optical \ion{Fe}{ii} lines, reminiscent of the Weigelt blobs in the local LBV star Eta Carinae. 
   Spectra and images of Godzilla and two faint adjacent images, and the detection of a low-surface brightness foreground galaxy in the NIRCam data, support the interpretation that Godzilla is an extremely magnified object due to the alignment with lensing caustics. We find that Godzilla is part of a previously identified clump, comprising $\sim10 - 25$ \% of it, with magnifications in the range of $\approx600-25,000$ depending on the models and images in comparison.
   The unique \ion{O}{i} source in Godzilla is well explained by a non-erupting LBV accompanied by a hotter companion and/or gas condensations exposed to more intense radiation compared to the Weigelt blobs. 
   If Godzilla is confirmed to contain an LBV star, it expands the distance to the furthest known LBV from a dozen Mpc to several Gpc.}
   {}


   \maketitle
%

\section{Introduction} \label{sec:introduction}
Gravitational lensing magnifies distant objects. 
For a sufficiently small source (e.g. a single star), the magnification from a smooth galaxy cluster-scale lens can reach extreme factors \citep[$10^3 - 10^6$,][]{miralda-escude1991}.
In realistic cluster-scale lenses, theoretical work has found that ``microlenses", small-scale masses (e.g. stars) within the lensing cluster, perturb the lensing potential near the primary caustics, creating a web of micro-caustics. These microcaustics cause the total magnification of the background lensed star to fluctuate, and also tend to limit the maximum magnification achievable \citep{venumadhav2017,diego2018,diego2019}.

The first lensed stars at cosmological distances ($z\sim 1-1.5$) were discovered as transients in Hubble Space Telescope (HST) imaging, their magnifications fluctuating as a result of microlensing \citep{kelly2018,rodney2018,kaurov2019,chen2019}. 
Since these initial discoveries, many more lensed stars have been found using this transient method \citep[e.g.,][]{kelly2022,fudamoto2024}.
Additionally, several lensed stars have been identified from their proximity to the lensing critical curve \citep[e.g.,][]{meena2023,diego2023a,diego2023}, including the most distant lensed star yet discovered at $z\sim 6$ \citep{welch2022,welch2022a}.

Thus far, spectroscopic studies of lensed stars have proven difficult, owing to their transient nature or faintness. 
\cite{furtak2024} present the James Webb Space Telescope (JWST) NIRSpec prism spectroscopy of a lensed star at $z=4.76$, however they only detect stellar continuum, making it difficult to draw precise conclusions on the nature of the star. 

In this paper, we study a lensed star candidate in the Sunburst Arc galaxy at $z=2.37$.
The Sunburst Arc is the brightest known strongly lensed galaxy at optical wavelengths, with an integrated magnitude m$_{\rm AB} \lesssim 18$ \citep{dahle2016} and dozens of highly magnified and multiply imaged star-forming regions \citep{sharon2022}. The Sunburst Arc earns its name from the strong ``direct escape'' Lyman-$\alpha$ (\lya) emission \citep{rivera-thorsen2017} and high fraction of Lyman Continuum (LyC) radiation escaping along the line of sight (LOS) from a single star-forming region \citep{rivera-thorsen2019}. 

As a result of strong gravitational lensing, unresolved features in the Sunburst Arc (clumps or knots) are replicated multiple times along the arc. For example, the LyC-emitting clump (Sunburst LCE) shows an unusually extensive collection of 12 copied images. One unique clump as bright as Sunburst LCE, however, does not appear to have a counterpart anywhere else (\autoref{fig:overview}). \cite{vanzella2020} found that this singly-imaged source in the northwest arc also exhibits highly unusual rest-ultraviolet (UV) features, such as \ion{Fe}{ii} emission lines excited by \lya\ pumping \citep[Bowen fluorescence,][]{bowen1934}, and has very high electron density $n_{\rm e} \gtrsim 10^6\: {\rm cm}^{-3}$ probed by carbon and silicon emission doublets. \cite{vanzella2020} first denoted this object as ``Tr'', meaning transient. \cite{diego2022} followed this notation while dubbing the object ``Godzilla'', and \cite{sharon2022} referred to it as ``a discrepant clump''. Recently, \cite{pascale2024} also adopted the name Godzilla, so we use this term in this paper for the sake of a homogeneous terminology. For arcs and the IDs of the lensed images of clumps, we adopt the terminology of \cite{sharon2022}.

Godzilla is believed to be an object residing in the Sunburst Arc galaxy, as its spectroscopic redshift is the same as the Sunburst Arc  ($z \approx 2.37$). The absence of any counterimages of Godzilla can be explained by an additional microlensing effect \citep{weisenbach2024} caused by a small object in front of Godzilla, extremely magnifying it only at that position \citep{diego2022,sharon2022}. \cite{pascale2024} argues that Godzilla is a young, massive star cluster from spectral energy distribution (SED) fitting, but the absence of Godzilla counterimages is not clearly explained in that work.

An earlier explanation of Godzilla as a transient such as a supernova (SN) \citep{vanzella2020} has since been ruled out \cite{diego2022} and \cite{sharon2022}. \citeauthor{diego2022} argued that Godzilla is likely a long-duration or stable source rather than a typical SN based on that it has stayed bright for over 5 years in the observer's frame (from 2016 to 2021). It far surpasses typical SN durations of 1 to several months although \cite{vanzella2020} suggested at least 1 year of durations for a SN interacting with its circumstellar medium (CSM). \citeauthor{diego2022} also mentions that archival imaging from 2014 \citep{dahle2016} shows Godzilla as a bright, unresolved source. \citeauthor{sharon2022} demonstrated that Godzilla did not clearly fade from February 2018 to December 2020 in HST imaging (Figure 9 therein). Depending on which observation is counted, Godzilla maintained its luminosity for 1 (counted from 2018) to 1.93 (counted from 2014) years in the rest frame, correcting for cosmological time dilation. The SN scenario also contradicts the maximum time delay predicted from the lens model by \cite{sharon2022}. Relative time delays between the north arc and the northwest and west arcs are less than a year \citep[][Figure 8]{sharon2022}, so counterimages of Godzilla should have been observed in other arcs if Godzilla were a SN.

Previous studies on Godzilla gave special attention to $\eta$ Carinae ($\eta$ Car), a Luminous Blue Variable (LBV) star in the Milky Way, as a possible analog. LBV is an umbrella term that includes previously defined Hubble-Sandage Variables, S Dor Variables, P Cyg and $\eta$ Car type stars as sub-types, and specifically excludes Wolf-Rayet stars and blue supergiants \citep{conti1984, weis2020}. LBVs are massive evolved stars \citep[see their Figure 3]{weis2020} often surrounded by circumstellar nebulae as a consequence of their high mass loss rate \citep{weis2020}. These hot, unstable stars occupy an inclined instability strip of $-11 \leq \rm M_{bol} \leq -9$ and $14,000 \leq T_{\rm eff} \leq 35,000$ K in the Hertzsprung-Russell (HR) diagram when they are quiescent, i.e. not erupting \citep[S~Dor cycle,][]{wolf1989}. In some cases, they undergo ``giant eruptions'' where the brightness increases by several magnitudes, as spectacularly observed for $\eta$ Car in the year 1843 \citep[``The Great Eruption'',][]{humphreys1999} or in P Cygni in the 17th century \citep{degroot1988}. 
During the Great Eruption, $\eta$ Car brightened to $\rm m_V \approx -1$ (currently $\rm m_V \approx 4.0$\footnote{Martin, J., 2024, Observations from the AAVSO International Database, https://www.aavso.org}) and ejected at least 10~$\rm M_{\odot}$, forming the Homunculus nebula \citep[Ch.~7]{davidson2012}. \cite{diego2022} demonstrated that the observed brightness of Godzilla can be achieved assuming a magnification factor of $\mu \approx 7000$ and an intrinsic brightness similar to that of $\eta$ Car during the Great Eruption.

\cite{vanzella2020} also pointed out that the \lya-pumped \ion{Fe}{ii} \lam1914 detected in Godzilla has been observed in the Weigelt blobs in $\eta$ Car \citep{johansson2005}. \cite{weigelt1986b} first discovered these three star-like gas condensations with diameters of < \ang{;;0.03} ($\sim 70$ AU for a distance of 2300 pc) located within \ang{;;0.1}- \ang{;;0.2} from $\eta$ Car using speckle interferometry. In 1995, HST resolved them in spectroscopy \citep[Ch. 1.4]{davidson2012} and revealed that they are slowly moving with a line-of-sight velocity of $\sim 40$ \kms, have densities of $10^7 - 10^8 \:\rm cm^{-3}$, and temperatures of 6,000 - 7,000 K \citep[Ch. 5]{davidson2012}. HST spectroscopy also revealed the 5.54-year spectroscopic cycle of $\eta$ Car \citep{damineli1996}, where high excitation lines weaken when the hotter secondary star ($40-50$ M$_\odot$, $\sim 4 \times 10^5 L_\odot$, $T_{\rm eff} \approx 39,000$ K; \citealt{mehner2010}) hides behind the more massive, cooler primary star ($\gtrsim 90$ M$_\odot$, $\sim 10^{6.7} L_\odot$, $T_{\rm eff} \sim 20,000$ K; \citealt[1.3.1]{davidson2012}). Along with high excitation lines, \lya-pumped and \lyb-pumped lines such as \ion{Fe}{ii} \lam8453, \ion{Fe}{ii} \lam8490 and \ion{O}{i} \lam8449 strengthen when the Weigelt blobs are exposed to the hotter secondary star \citep[Figure 5.4]{davidson2012}. 

The origin of the Weigelt blobs is the subject of ongoing research. It has been previously believed that the Weigelt blobs were formed during a brightening in 1941 
\citep{davidson2012, abraham2014}. \cite{abraham2020} use Atacama Large Millimeter Array (ALMA) observations to better constrain the proper motion (and hence the formation time) of the Weigelt blobs. They find that the three blobs are formed at different times (see their Fig.16), but in each case the blob formation coincides with epochs of minimum intensity in high-ionisation lines in the spectrum of $\eta$ Car. These epochs are also thought to coincide with the periastron of the hotter companion. There do not seem to be particularly noticeable enhancements of the overall brightness of eta Car during these epochs \citep[see e.g. the V-band/visual light curve in Fig. 3 of ][]{fernandez-lajus2009}.  

Detecting extragalactic LBVs in spatially unresolved observations is highly challenging \citep[e.g.][]{guseva2023}. First, it is hard to capture the LBV phase of a star due to the short lifespan of this phase \citep[$10^4-10^5$ years assuming single star evolution]{herrero2010}. Also, LBVs are not discernible from other bright hot stars during their quiescent phase, which can last for decades or centuries \citep{wofford2020a}. When they are not erupting, the S Dor variability is the only characteristic that distinguish LBVs from other massive, evolved stars \citep{weis2020}. For these reasons, only a handful of LBVs have been observed, and the farthest confirmed LBV is in DDO 68 which is 12.75 Mpc away \citep{makarov2017}. If Godzilla is really an LBV, then this discovery extends the farthest known individual LBV star from one at a dozen Mpc to several Gpc ($z = 2.37$). However, many important traits of LBVs, such as broad hydrogen and often helium lines associated with P Cygni profiles \citep{guseva2023}, emission line nebulae and dust nebulae often surrounding LBVs, are not possible to examine through ground-based telescopes at this redshift. Moreover, it was not possible to separate a pure spectrum of Godzilla uncontaminated by other features in previous ground-based slit spectroscopy. Thus, previous suggested explanations of Godzilla's observed properties are based on the detection of a few exceptional emission lines or luminosity constraints from lens models. In this paper, we present  JWST/NIRCam imaging and NIRSpec IFU observations of Godzilla for the first time. We have extracted the spatially resolved rest-optical to rest-near infrared (NIR) spectrum of Godzilla. Using imaging and spectroscopic data of unprecedented quality, we try to unveil the true nature of Godzilla.

The rest of this paper is structured as follows: in Section~\ref{sec:obs} we describe the observations and data reduction. In Section~\ref{sec:methods} we explain how we extracted spectra from Godzilla and four regions surrounding it. We also describe how we measured the kinematics and fluxes of emission lines, and corrected measured flux for dust reddening. In Section~\ref{sec:results}, we describe main results, including the identification of almost 60 emission lines, kinematics and gas properties of Godzilla, and the detection of \lyb-pumped \ion{O}{i} \lam8449 and \lya-pumped iron lines (Section~\ref{ssec:bowen}). In Section~\ref{sec:discussion}, we discuss various aspects of Godzilla. We first take a look at the NIRCam image and revisit SN and microlensing scenarios (Section~\ref{ssec:lensing}). Then we identify counterimages including Godzilla, and derive magnificaion of Godzilla from them (Section~\ref{ssec:magnification}). To understand the \ion{O}{i} \lam8449 source sitting in Godzilla, we investigate spatial variations in the kinematics, dust, and gas properties by analyzing the surrounding regions (Section~\ref{sssec:spatial_analysis}). We go through \ion{O}{i} \lam8449 source candidates and exclude other scenarios (Section~\ref{sssec:excluded_candidates}), showing that an LBV possibly with a hot companion analogous to $\eta$ Car best explains the \ion{O}{i} \lam8449 source (Section~\ref{sssec:lbv_scenario}).
 

This paper assumes a flat $\Lambda$CDM cosmology with parameters $\Omega_{\Lambda} = 0.7$, $\Omega_{m}=0.3$, and $H_0 = 70$ km s$^{-1}$ Mpc$^{-1}$. With this cosmology, the redshift of the source, $z=2.37$, corresponds to 2.72 Gyr after the Big Bang.
   
\section{Observations} \label{sec:obs}

\begin{figure*}
    \centering
    \includegraphics[width=.9\textwidth]{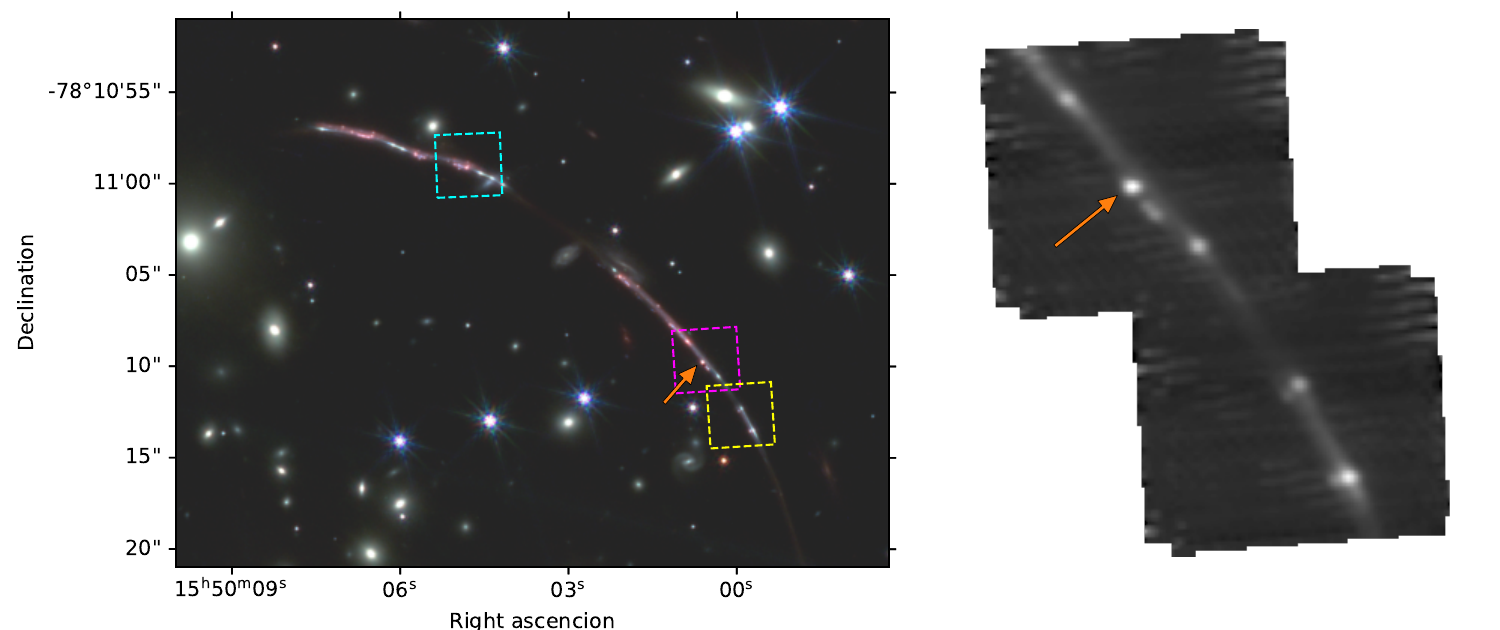}
    \caption{(left) Overview of the north and northwest arc with the NIRSpec IFU pointings overlaid. Cyan, magenta, and yellow squares depict pointing 1, 2, and 3, respectively. We used NIRCam F115W, F200W and F444W filters for R,G,B, respectively. Godzilla in pointing 2 is marked with an orange arrow. (right) Pointing 2 and 3 showing Godzilla marked with an orange arrow. The images are oriented such that N is up, E is left; each combined NIRSpec pointing is approximately $3\farcs5$ on each side.}\label{fig:overview}
\end{figure*}

\begin{figure*}
   \centering
\includegraphics[width=\textwidth]{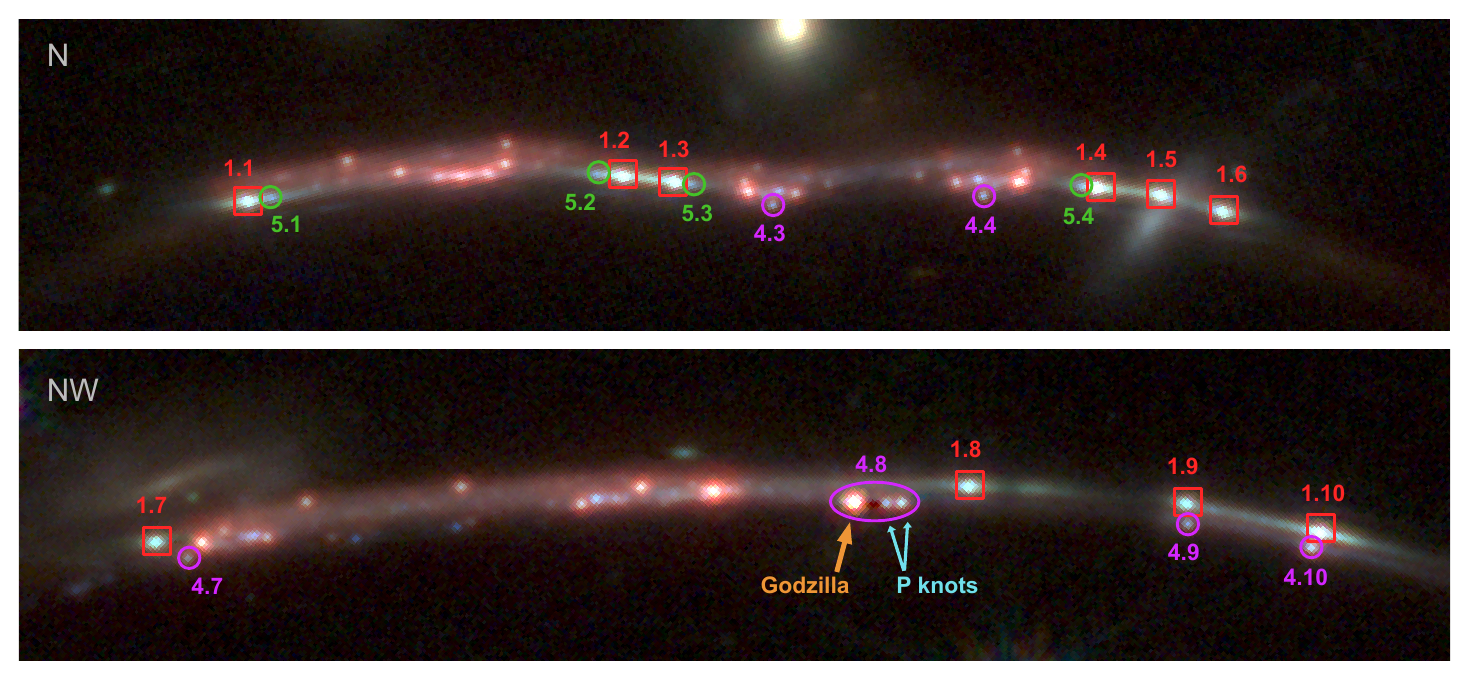}
   \caption{The composite of NIRCam F115W, F200W and F444W images of the north and the northwest arc of the Sunburst Arc. Images 3 - 10 of clump 4 are marked with purple circles, while images 1 - 10 of the LyC leaking clump (clump 1) are marked with red squares. In image 8 of clump 4 (4.8), we observe Godzilla and the P knots (the two small clumps). Image 1 - 4 of clump 5 are marked with green circles. These 4 images and image 4.7 (t1 - t5 in \cite{diego2022}) have been suggested as candidates for the counterimages of Godzilla by \cite{diego2022}.}     \label{fig:nircam_image}%
    \end{figure*}

JWST observed the Sunburst Arc with NIRCam imaging and NIRSpec integral field spectroscopy (IFS) in the interval April 4, 2023 to April 10, 2023 (JWST Cycle 1, GO-2555, PI: Rivera-Thorsen). Imaging was done in each of the NIRCam filters F115W, F150W, F200W, F277W, F356W, and F444W. Two NIRSpec grating and filter combinations, G140H/F100LP and G235H/F170LP, were used to cover a rest-frame wavelength range of
\(0.29 \mu \rm{m} \le \lambda_0 \le 0.56 \mu \rm{m}\) and \(0.49 \mu \rm{m} \le \lambda_0 \le 0.94 \mu \rm{m}\), respectively. Of the three on-target NIRSpec pointings, Godzilla is found in the pointing labeled 2 (the magenta square in \autoref{fig:overview}, indicated with an orange arrow). We refer to a companion paper by Rivera-Thorsen et al. (in prep.) for a full description of observations and data reduction.


This work also used observed-frame optical (rest-UV), ground-based spectra collected with the Magellan Echellette (MagE) spectrograph on the Magellan-I Baade Telescope of the Las Campanas Observatory in Chile. This spectrum is the slit M3 pointing displayed in Figure 1 of \citet{mainali2022} that covers Godzilla and two smaller, adjacent knots (the ``P knots'', following the nomenclature of \cite{diego2022}) seen in \autoref{fig:nircam_image}. To summarize, we positioned slit M3 on the arc by comparing the MagE slit viewing camera and HST images, and reduced the data as described in \citet{rigby2018}. For more in-depth descriptions of the MagE observations and the reduction process, we refer to \cite{owens2024} for this pointing specifically, and Rigby et al. (in prep.) will present a complete observation log for each pointing of the parent MagE observation program.

\section{Methods} \label{sec:methods}
   \begin{figure*}
   \centering
   \includegraphics[width=\textwidth]{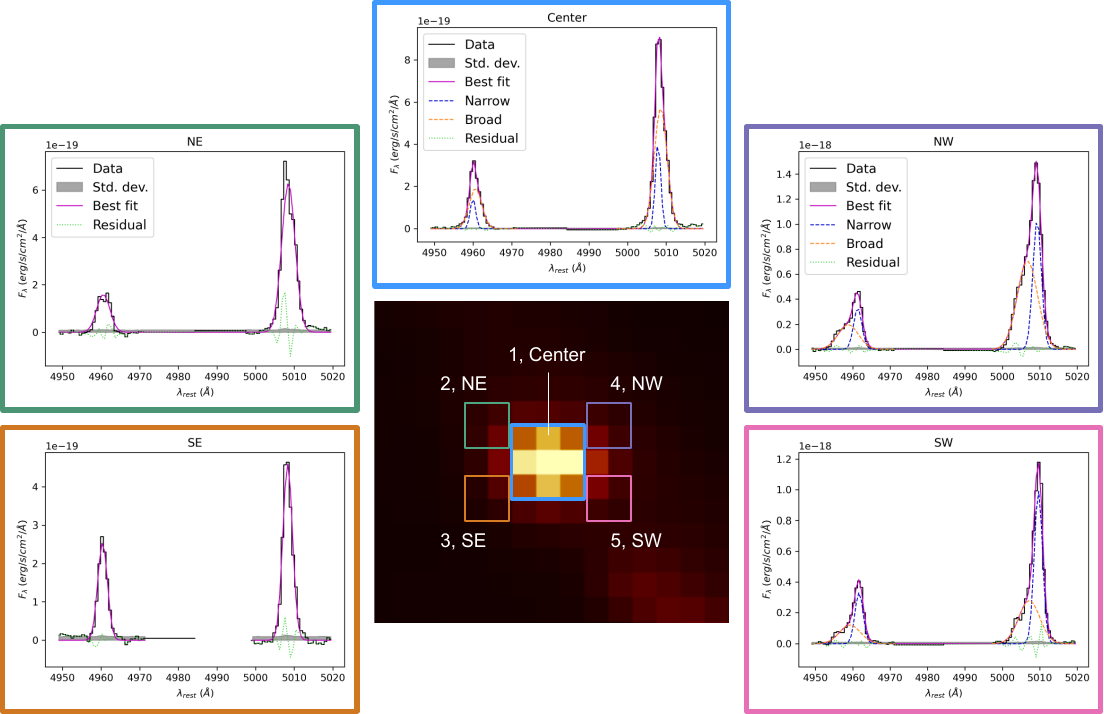}
 \caption{The 5 regions where spectra have been extracted and their [\ion{O}{iii}]$\lambda\lambda$4960,5008 doublet line profiles in the G140H/F100LP grating. The center figure shows the wavelength-axis median image of the NIRSpec G140H/F100LP grating at the area of Godzilla. Two component (Center, NW, SW) and one component (NE, SE) Gaussian fits are also shown.} \label{fig:five_regions}
    \end{figure*}

    \begin{figure*}
   \centering
   \includegraphics[width=\textwidth]{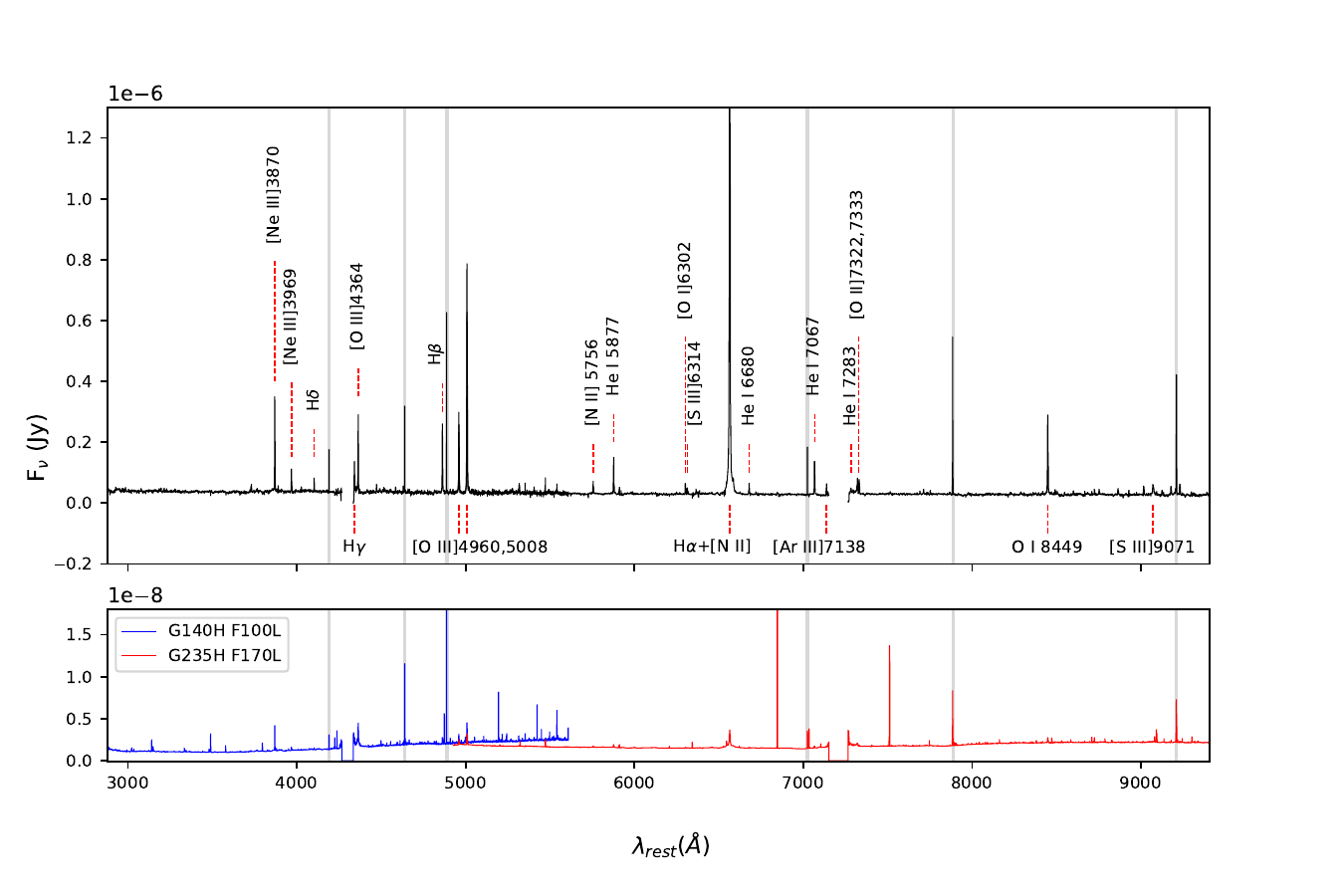}
   \caption{(top) One dimensional spectrum extracted from the Center region, labeled with prominent lines. (bottom) Corrected error from G140H/F100LP and G235H/F170LP gratings shown in blue and red, respectively. Noise pixels are marked with grey shades. Both flux density and errors from the two gratings are overlapped in $\sim 4925 - 5610$ {\AA}. See \autoref{fig:spectra_1} and \autoref{fig:spectra_2} for more detailed features.}
              \label{fig:full_spectrum}%
    \end{figure*}

   \begin{figure}
   \centering
\includegraphics[width=\columnwidth]{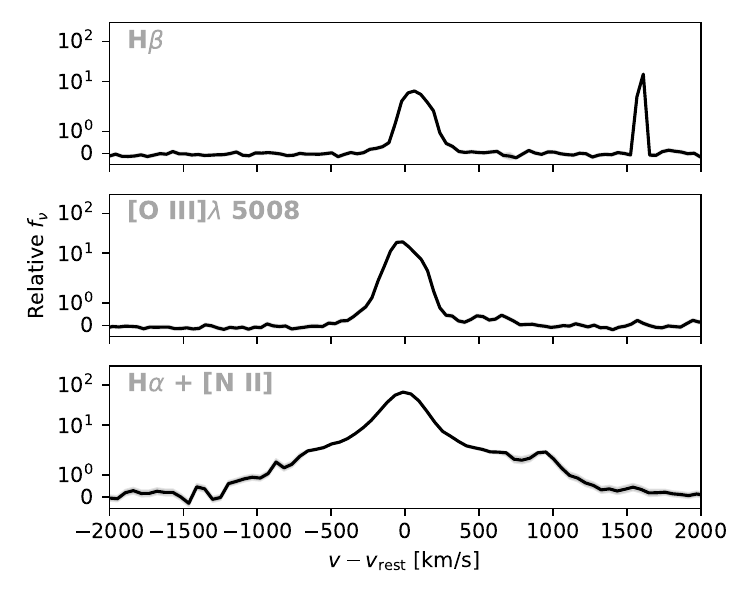}
   \caption{Line profiles of H$\beta$, [\ion{O}{iii}]5008, and H$\alpha$+[\ion{N}{ii}] of the Center region, highlighting the broad component present in H$\alpha$ but not the other lines. The y-axis scale is linear between -1 and 1, and logarithmic outside this interval. The velocity scale in the lower panel is centered around H$\alpha$.}
              \label{fig:broad_ha_symlog}%
    \end{figure}

   \begin{figure*}
   \centering
   \includegraphics[width=\textwidth]{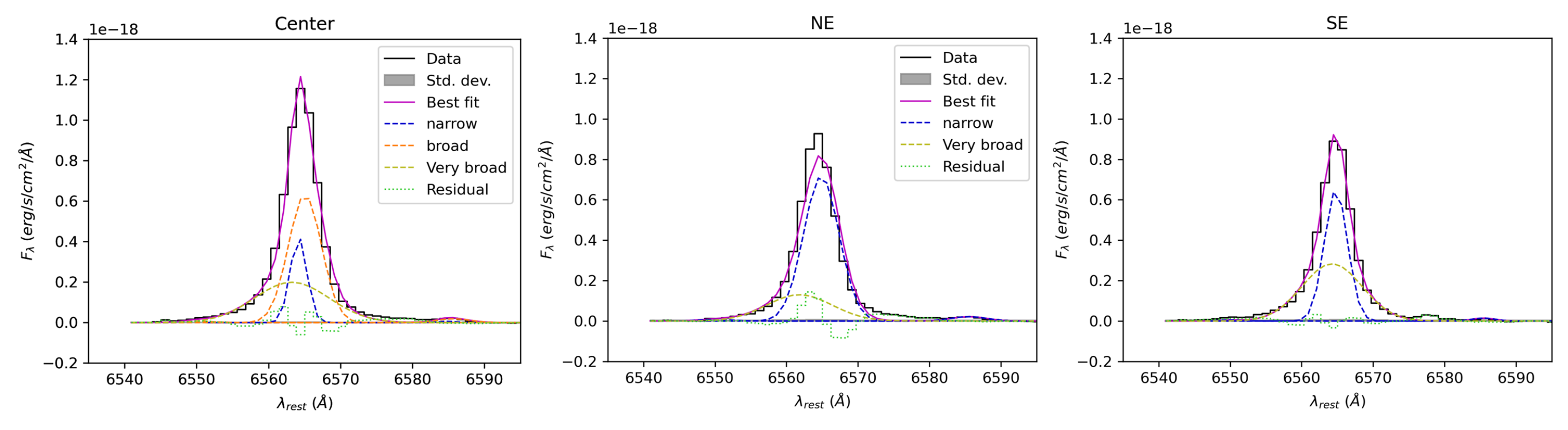}
 \caption{Fitting result of H$\alpha$ including a very broad component additional to default two component (Center) and one component (NE, SE) Gaussian profiles.}
              \label{fig:Ha_very_broad}%
    \end{figure*}

\subsection{Extraction of one dimensional spectra}\label{ssec:extraction}

We extracted spectra from the brightest center of Godzilla (Center), four regions surrounding it (NE, SE, NW, SW) to investigate spatial variations of gas properties (\autoref{fig:five_regions}), and the two P knots (named P1 and P2 for the left and right knot, respectively) for visual inspection. The spectra were extracted as a continuum-weighted spaxel average. The extracted spectrum of Center appears in the top panel of \autoref{fig:full_spectrum}, with prominent lines labeled. The MAST instrument pipeline appears to underestimate errors, considering that the pipeline errors are smaller than the standard deviation of line-masked continuum regions in the spectrum itself. To correct this, we added the standard deviation of an empty region of sky in quadrature with the pipeline's error estimate. See Rivera-Thorsen et al. (in prep.) for a more in-depth explanation. At the bottom panel of \autoref{fig:full_spectrum}, we show corrected uncertainties in the flux density for the G140H/F100LP and G235H/F170LP gratings in blue and red colors, respectively. 

Before measuring fluxes, we subtracted the continua from the spectra. We first calculated the running-median and standard deviation ($\sigma$) of the spectrum with a rest-frame 100 \AA\ wide moving window. After removing data outside 3$\sigma$, we re-calculated the running median at the moving window, and subtracted that from the original spectrum.

\subsection{Flux measurement} \label{ssec:flux}
\begin{table*}[]
\caption{Dust-corrected line fluxes for the Center G140H/F100LP spectrum. 
\label{tab:blue_flux}}
\centering
\begin{tabular}{llcccccl}
\toprule
   &  &          & \multicolumn{4}{c}{F($\lambda$)/F(H$\beta$) [\%]\tablefootmark{a}}                 &        \\
   \cmidrule(lr){4-7}
\# &
  \multicolumn{1}{c}{Line} &
  \multicolumn{1}{c}{$\rm\lambda_{vac}$ [\AA]\,\tablefootmark{b}} &
  \multicolumn{1}{c}{Narrow} &
  \multicolumn{1}{c}{Broad} &
  \multicolumn{1}{c}{Total} &
  \multicolumn{1}{c}{Total (ext. corr.)} &
  \multicolumn{1}{c}{S/N\, \tablefootmark{c}} \\ \midrule
1  & H14               & 3723.00 & 2.8$\pm$0.7   & 0.5$\pm$0.9   & 3.3$\pm$0.5   & 5.6$\pm$0.8   & 7   \\
2  & {[}O II{]} 3727   & 3727.09 & 4.1$\pm$1.6   & 7.0$\pm$2.9   & 11.2$\pm$2.4  & 19$\pm$4      & 4.7   \\
3  & {[}O II{]} 3730   & 3729.88 & 8.3$\pm$2.1   & 4.0$\pm$3.4   & 12.2$\pm$2.4  & 21$\pm$4      & 5.0   \\
4  & {[}Ne III{]} 3870 & 3869.86 & 95.1$\pm$1.9  & 47.4$\pm$1.3  & 142.5$\pm$1.4 & 222.7$\pm$2.1 & 105    \\
5  & H8                & 3890.17 & 6.5$\pm$0.4   & 0.0$\pm$0.4   & 6.5$\pm$0.4   & 10.0$\pm$0.6  & 17  \\
6  & {[}Ne III{]} 3969 & 3968.59 & 19.4$\pm$0.8  & 16.4$\pm$1.2  & 35.8$\pm$0.6  & 53.3$\pm$0.9  & 58  \\
7  & H7                & 3971.20 & 4.6$\pm$0.6   & 1.5$\pm$0.9   & 6.1$\pm$0.6   & 9.0$\pm$0.9   & 10  \\
8  & He I 4027         & 4027.33 & 4.5$\pm$0.5   & 0.2$\pm$0.5   & 4.7$\pm$0.4   & 6.7$\pm$0.6   & 12  \\
9  & H-$\delta$        & 4102.90 & 10.6$\pm$0.6  & 7.5$\pm$0.9   & 18.0$\pm$0.5  & 25.1$\pm$0.7  & 35  \\
10 & H-$\gamma$        & 4341.69 & 16.0$\pm$1.0  & 31.7$\pm$1.4  & 47.7$\pm$0.9  & 59.3$\pm$1.1  & 56  \\
11 & {[}O III{]} 4364  & 4364.44 & 65.1$\pm$1.4  & 14.6$\pm$1.6  & 79.6$\pm$1.1  & 98.0$\pm$1.3  & 74  \\
12 & Fe II 4418        & 4418.07 & 2.3$\pm$0.5   & 0.4$\pm$0.6   & 2.7$\pm$0.4   & 3.2$\pm$0.5   & 6   \\
13 & {[}Fe II{]} 4453  & 4453.35 & 1.6$\pm$0.4   & 0.00$\pm$0.14 & 1.6$\pm$0.4   & 1.8$\pm$0.4   & 4.4   \\
14 & He I 4473         & 4472.73 & 6.4$\pm$0.6   & 1.9$\pm$0.9   & 8.3$\pm$0.6   & 9.8$\pm$0.7   & 15  \\
15 & Fe II 4490        & 4490.44 & 2.4$\pm$0.4   & 0.0$\pm$0.5   & 2.4$\pm$0.5   & 2.8$\pm$0.5   & 5   \\
16 & Fe II 4493        & 4492.67 & 0.00$\pm$0.2  & 4.5$\pm$0.6   & 4.5$\pm$0.6   & 5.2$\pm$0.7   & 8   \\
17 & Fe II 4557        & 4557.17 & 2.2$\pm$0.5   & 0.8$\pm$0.8   & 3.0$\pm$0.5   & 3.4$\pm$0.6   & 6   \\
18 & Fe II 4585        & 4585.12 & 4.7$\pm$0.6   & 0.8$\pm$0.8   & 5.5$\pm$0.5   & 6.1$\pm$0.6   & 10   \\
19 & Fe II 4631        & 4630.64 & 4.6$\pm$0.4   & 0.00$\pm$0.19 & 4.6$\pm$0.4   & 5.0$\pm$0.4   & 12  \\
20 & {[}Fe III{]} 4659 & 4659.35 & 2.9$\pm$0.6   & 0.7$\pm$0.8   & 3.6$\pm$0.5   & 3.9$\pm$0.6   & 7   \\
21 & H-$\beta$\,\tablefootmark{d}        & 4862.69 & 25.8$\pm$0.7  & 74.2$\pm$1.1  & 100.0$\pm$0.7 & 100.0$\pm$0.7 & 143 \\
22 & {[}O III{]} 4960 & 4960.30 & 26.7$\pm$0.8 & 81.9$\pm$1.1  & 108.5$\pm$0.7 & 104.5$\pm$0.7 & 146 \\
23 & {[}O III{]} 5008 & 5008.24 & 76$\pm$1 & 250$\pm$2 & 326$\pm$1 & 308$\pm$1 & 344  \\
24 & Fe II 5017        & 5016.92 & 1.5$\pm$0.7   & 5.3$\pm$1.1   & 6.8$\pm$0.7   & 6.4$\pm$0.6   & 10   \\
25 & Fe II 5020        & 5019.84 & 1.8$\pm$0.6   & 2.8$\pm$1.0   & 4.6$\pm$0.7   & 4.4$\pm$0.7   & 7  \\
 \bottomrule
\end{tabular}
\tablefoot{\\
\tablefoottext{a}{Line fluxes are normalized by the total best-fit flux of H$\beta$, but not including its uncertainty.}\\
\tablefoottext{b}{Wavelengths from the NIST Atomic Spectral Database \citep{kramida1999}.}\\
\tablefoottext{c}{Signal-to-noise of the total, measured (uncorrected) line flux; rounded off to nearest integer unless where lower than 5.}\\
\tablefoottext{d}{F(H$\beta$)$_{\text{obs}}=3.216\pm0.022 \times 10^{-18}$, F(H$\beta$)$_{\text{corr}}=2.156\pm0.015 \times 10^{-17}$ (erg/s/cm$^2$).}
}
\end{table*}

\begin{table*}[]
\centering
\caption{Measured and dust corrected line fluxes for the Center G235H/F170LP spectrum.
\label{tab:red_flux}}
\begin{tabular}{lllccccl}
\toprule
   & \multicolumn{1}{l}{} &          & \multicolumn{4}{c}{F($\lambda$)/F(H$\beta$)   [\%]}                                   &        \\
   \cmidrule(lr){4-7}
\# &
  Line &
  \multicolumn{1}{c}{$\rm\lambda_{vac}$ [\AA]} &
  \multicolumn{1}{c}{Narrow} &
  \multicolumn{1}{c}{Broad} &
  \multicolumn{1}{c}{Total} &
  \multicolumn{1}{c}{Total   (ext. corr.)} &
  \multicolumn{1}{c}{S/N} \\ \midrule
22 & {[}O III{]} 4960  & 4960.30 & 39.3$\pm$1.5    & 77.1$\pm$1.8   & 116.4$\pm$0.9  & 112.1$\pm$0.8  & 137 \\
23 & {[}O III{]} 5008  & 5008.24 & 102.1$\pm$1.8   & 220.2$\pm$2.1  & 322.4$\pm$1.0  & 305.0$\pm$1.0  & 316 \\
24 & Fe II 5017        & 5016.92 & 1.1$\pm$1.5     & 6.1$\pm$2.3    & 7.3$\pm$1.1    & 6.9$\pm$1.0    & 7   \\
25 & Fe II 5020        & 5019.84 & 0.0$\pm$0.9     & 6.0$\pm$1.2    & 6.0$\pm$1.1    & 5.6$\pm$1.0    & 6   \\
26 & {[}N II{]} 5756   & 5756.19 & 10.4$\pm$0.7    & 2.2$\pm$1.0    & 12.7$\pm$0.5   & 9.31$\pm$0.35  & 27  \\
27 & He I 5877         & 5877.25 & 25.0$\pm$0.8    & 13.7$\pm$1.0   & 38.7$\pm$0.5   & 27.4$\pm$0.4   & 76  \\
28 & {[}O I{]} 6302    & 6302.05 & 7.1$\pm$0.5     & 3.4$\pm$0.7    & 10.5$\pm$0.4   & 6.69$\pm$0.24  & 28  \\
29 & {[}S III{]} 6314  & 6313.81 & 4.6$\pm$0.5     & 0.7$\pm$0.6    & 5.3$\pm$0.4    & 3.36$\pm$0.23  & 15  \\
30 & {[}O I{]} 6366    & 6365.54 & 2.9$\pm$0.5     & 1.5$\pm$0.7    & 4.4$\pm$0.4    & 2.76$\pm$0.24  & 11  \\
31 & Si II 6373        & 6373.13 & 0.6$\pm$0.5     & 2.1$\pm$0.7    & 2.7$\pm$0.4    & 1.66$\pm$0.23  & 7   \\
32 & {[}N II{]} 6550   & 6549.86 & 0.71$\pm$0.01  & 3.85$\pm$0.11  & 4.56$\pm$0.12  & 2.70$\pm$0.07  & 37  \\
33 & H-$\alpha$\,\tablefootmark{\dag}        & 6564.63 & 128$\pm$4       & 380$\pm$12     & 509$\pm$16     & 300$\pm$9      & 32  \\
   & H-$\alpha$\,\tablefootmark{\ddag}        & 6564.63 & --- &  ---  & 781$\pm$9  & ---  & ---  \\
34 & {[}N II{]} 6585   & 6585.28 & 2.10$\pm$0.04   & 11.40$\pm$0.33 & 13.5$\pm$0.4   & 7.93$\pm$0.22  & 37  \\
35 & He I 6680         & 6679.99 & 6.5$\pm$0.5     & 2.2$\pm$0.6    & 8.74$\pm$0.35  & 5.01$\pm$0.20  & 25  \\
36 & {[}S II{]} 6718  & 6718.29 & $(4\pm2)\times10^{-4}$ & 1.6$\pm$0.3       & 1.6$\pm$0.3       & 0.9$\pm$0.2 & 4.7  \\
37 & {[}S II{]} 6733   & 6732.67 & 0.66$\pm$0.23   & 0.00$\pm$0.04  & 0.66$\pm$0.23  & 0.37$\pm$0.13  & 2.9   \\
38 & He I 7067         & 7067.14 & 17.1$\pm$0.4    & 12.2$\pm$0.6   & 29.27$\pm$0.35 & 15.27$\pm$0.18 & 83  \\
39 & {[}Ar III{]} 7138 & 7137.76 & 5.4$\pm$0.4     & 3.0$\pm$0.6    & 8.4$\pm$0.4    & 4.29$\pm$0.18  & 23  \\
40 & He I 7283         & 7283.36 & 2.58$\pm$0.34   & 0.0$\pm$0.4    & 2.60$\pm$0.32  & 1.29$\pm$0.16  & 8   \\
41 & {[}O II{]} 7322   & 7322.01 & 7.7$\pm$0.5     & 2.3$\pm$0.6    & 10.0$\pm$0.4   & 4.94$\pm$0.18  & 27  \\
42 & {[}O II{]} 7333   & 7332.75 & 7.88$\pm$0.26   & 0.00$\pm$0.19  & 7.88$\pm$0.25  & 3.88$\pm$0.12  & 32  \\
43 & {[}Ar III{]} 7753 & 7753.19 & 1.60$\pm$0.31   & 0.4$\pm$0.4    & 1.98$\pm$0.28  & 0.90$\pm$0.13  & 7   \\
   & O I 7776          & 7776.31 & 0.04$\pm$0.13   & 0.00$\pm$0.05  & 0.04$\pm$0.14  & 0.02$\pm$0.06  & 0.3   \\
44 & O I 7990          & 7989.17 & 0.60$\pm$0.22   & 0.00$\pm$0.10  & 0.60$\pm$0.22  & 0.26$\pm$0.10  & 2.7   \\
45 & Pa17              & 8469.58 & 0.64$\pm$0.24   & 0.00$\pm$0.20  & 0.64$\pm$0.23  & 0.25$\pm$0.09  & 2.8   \\
46 & O I 8449          & 8448.68 & 26.8$\pm$0.4    & 25.0$\pm$0.6   & 51.8$\pm$0.4   & 20.75$\pm$0.15 & 139 \\
   & O I 8449\tablefootmark{*}      & 8448.68  &  --- & --- &47.8$\pm$0.4        &  ---  & 134 \\
   & Fe II 8453       & 8453.34 & $(3 \pm 6) \times 10^{-5}$       & $(3\pm 2)\times10^{-4}$ & $(3\pm 2)\times10^{-4}$ & $(1.3\pm 0.7)\times10^{-4}$ & 1.8   \\
& Fe II 8453\tablefootmark{*}          & 8453.34 & --- & --- & 3.02$\pm$0.26 & --- & 12   \\
47 & Fe II 8490        & 8492.44 & 1.71$\pm$0.34   & 0.8$\pm$0.5    & 2.47$\pm$0.33  & 0.98$\pm$0.13  & 8   \\
   & Fe II 8490\tablefootmark{*} & 8492.44 & --- & --- & 2.58$\pm$0.26  & --- & 10   \\
48 & Pa16              & 8504.82 & 0.76$\pm$0.32   & 1.0$\pm$0.5    & 1.74$\pm$0.33  & 0.69$\pm$0.13  & 5   \\
49 & Pa15              & 8547.73 & 0.97$\pm$0.28   & 0.2$\pm$0.4    & 1.14$\pm$0.28  & 0.45$\pm$0.11  & 4.1   \\
50 & Pa14              & 8600.75 & 1.63$\pm$0.25   & 0.00$\pm$0.25  & 1.63$\pm$0.24  & 0.64$\pm$0.10  & 7   \\
51 & Pa13              & 8667.40 & 0.60$\pm$0.24   & 0.00$\pm$0.24  & 0.60$\pm$0.24  & 0.23$\pm$0.09  & 2.5   \\
52 & Pa12              & 8752.88 & 1.41$\pm$0.32   & 1.8$\pm$0.5    & 3.19$\pm$0.32  & 1.22$\pm$0.12  & 10   \\
53 & Pa11              & 8865.22 & 1.72$\pm$0.31   & 2.4$\pm$0.5    & 4.07$\pm$0.31  & 1.53$\pm$0.12  & 13  \\
54 & Pa10              & 9017.38 & 2.50$\pm$0.30   & 2.2$\pm$0.5    & 4.72$\pm$0.30  & 1.74$\pm$0.11  & 16  \\
55 & {[}S III{]} 9071 & 9071.09 & 0.000$\pm$0.03     & 9.0$\pm$0.3       & 9.0$\pm$0.3       & 3.3$\pm$0.1  & 28 \\
56 & {[}Cl II{]} 9126  & 9126.10 & 1.32$\pm$0.25   & 0.4$\pm$0.4    & 1.67$\pm$0.27  & 0.61$\pm$0.10  & 6    \\
57 & Pa9               & 9231.55 & 3.32$\pm$0.29   & 2.7$\pm$0.4    & 6.07$\pm$0.30  & 2.17$\pm$0.11  & 20  \\
\bottomrule
\end{tabular}
\tablefoot{Column description and notes in \autoref{tab:blue_flux} also apply here.\\
\tablefoottext{\dag}{H$\alpha$ line fitted with a third, very broad component, but the Total flux reported here is not including that; see next row.}\\
\tablefoottext{\ddag}{Total flux of H$\alpha$ including the third, very broad, Gaussian component.}\\
\tablefoottext{*}{Narrow, fluorescence-pumped lines re-fitted with only one Gaussian component and relaxed line width. See more in Section~\ref{sssec:laser}.}
}
\end{table*}

At Center, NW region, and SW region, we measured the flux of each line using two-component (narrow and broad) Gaussian fitting. Instrumental resolving power was accounted for by using the official NIRSpec calibration files\footnote{\url{https://jwst-docs.stsci.edu/jwst-near-infrared-spectrograph/nirspec-instrumentation/nirspec-dispersers-and-filters}}, and treating the linearly interpolated line spread function (LSF) as a Gaussian and adding that and the intrinsic linewidth in quadrature in the model. We first measured redshift and line width by fitting the strong oxygen doublet [\ion{O}{iii}]\lam4960 and [\ion{O}{iii}]\lam5008 simultaneously using a two-component Gaussian profile (\autoref{fig:five_regions}). Then we measured the flux of other lines in the same way (two-component Gaussian fitting), forcing each line to have the same redshift and velocity line width as measured for the [\ion{O}{iii}]\lam\lam4960,5008 doublet at each grating. 
The error of the narrow, broad and total fluxes were separately estimated using a Monte Carlo sampling method. The flux measured from the unperturbed spectrum was taken as the nominal value. For each wave bin, we then drew 999 random samples from a normal distribution with $(\mu, \sigma)$ being the observed flux density and the uncertainty in that bin, respectively, creating 999 perturbed spectra.
We repeated fitting for the perturbed samples, and reported the standard deviation of the 1,000 measured fluxes as error. We measured the flux in the NE and SE regions through the same procedure using only one Gaussian component, as additional components did not improve the fit here from visual inspection.

We found that the [\ion{O}{ii}]\lam\lam3727,3729 doublet is distinctively red-shifted compared to other lines (\autoref{fig:spectra_1}). To accurately recover the flux and flux ratio of the [\ion{O}{ii}]\lam\lam3727,3729 doublet, we allowed the redshift of these lines to vary, but kept the distance between the two lines fixed. At Center, the narrow component of [\ion{O}{ii}]\lam\lam3727,3729 doublet is offset relative to [\ion{O}{iii}] (and to nearby higher-order Balmer lines) by rest-frame $1.73\pm0.6$ \AA\ ($140\pm50$ \kms). The broad component has a small offset consistent with 0 within the error bars. 

In addition to a narrow and broad component, we found that a third, very broad component was necessary to obtain a good fit of H$\alpha$ at the Center, NE, and SE regions. We found no sign that this extra component was needed to obtain good fits for H$\beta$, [\ion{O}{iii}]\lam\lam 4960,5008, or any other line. \autoref{fig:broad_ha_symlog} clearly shows the necessity of the very broad component in fitting H$\alpha$. The narrow and broad components of the [\ion{N}{ii}]\lam\lam6550, 6585 doublet were set to have fixed ratios of 1:2.96 \citep{tachiev2001}. In our spectra, there are a few more doublets arising from the same upper level, which thus have fixed flux ratio (e.g. [\ion{O}{iii}]\lam\lam4960,5008, [\ion{N}{iii}]\lam\lam3870,3969, [\ion{O}{i}]\lam\lam 6302,6366), but we do not fix the flux ratio of these doublets. 
The two-Gaussian model is a simplification of the true line shape, and we found that additionally locking the line ratios led to over-constrained and over-simplified models producing poor fits. Allowing the line ratio to deviate slightly from the fixed value led to more accurate and better constrained models.
The [\ion{N}{ii}]\lam\lam6550,6585 doublet is the only exception from this, since [\ion{N}{ii}]\lam6550 is completely buried in the very broad component of H$\alpha$ and there is a chance of overestimating this broad component if we do not force the [\ion{N}{ii}]\lam6550 flux to be $\sim 1/3$ of the [\ion{N}{ii}]\lam6585 flux. \autoref{fig:Ha_very_broad} shows the fits of H$\alpha$ and [\ion{N}{ii}]\lam\lam6550,6585 with the third, very broad H$\alpha$ component included. In the NW and SW regions, the fits of the very broad component of H$\alpha$ had fluxes consistent with 0, so we do not include this third component for H$\alpha$ in these regions.

\subsection{Dust reddening}
   \begin{figure}
   \centering
\includegraphics[width=\columnwidth]{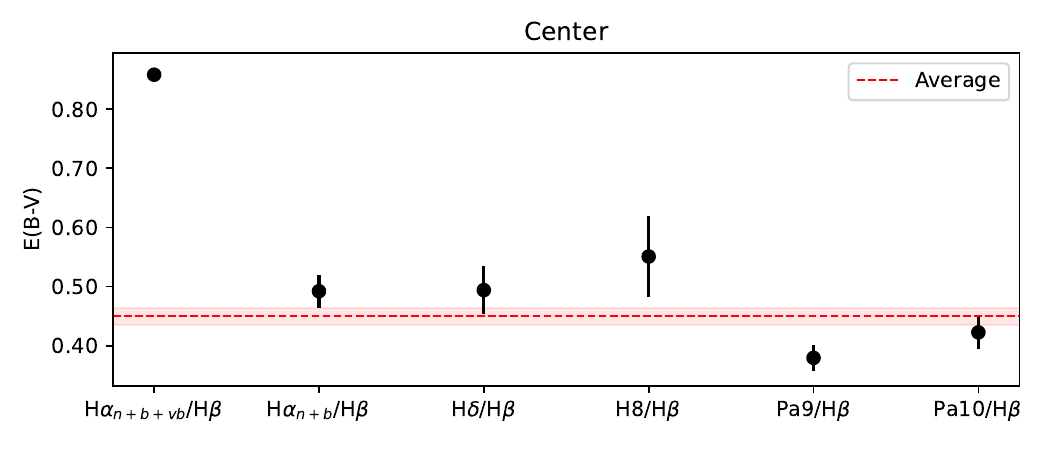}
 \caption{$E(B-V)$ as computed from the strongest Balmer and Paschen lines for the Center region. The value from H$\alpha$ is shown both including and excluding the very broad component. The average weighted by inverse of error is shown in red-dashed horizontal line with a shade representing 1 $\sigma$ uncertainty. Similar figures for the surrounding regions are shown in \autoref{fig:ebv_wings}.} \label{fig:ebv_Center}
    \end{figure}
    
We modeled dust attenuation using a starburst attenuation law \citep{calzetti2000}, assuming a standard $R_V = 4.05$. We determined $E(B-V)$ using H$\alpha$, H$\beta$, H$\delta$, H8, Pa9, and Pa10. H$\gamma$ has not been included since it partially lies on the detector gap, making its measured flux unreliable. We also do not include H7 as this line is blended with [\ion{Ne}{iii}]\lam3969. Intrinsic fluxes of Balmer and Paschen lines are calculated with PyNeb version 1.1.17 \citep{luridiana2015a} assuming $T_{\rm e} = 10^4$ K  and $n_{\rm e} = 10^3$ $\rm cm^{-3}$. The Balmer decrement can vary with density, but its effect is minimal. For example, at $T_{\rm e} = 10^4$ K, the H$\alpha$/H$\beta$ ratio is 2.86 at a density of $n_{\rm e} = 10^3$ $\rm cm^{-3}$ and 2.81 at $n_{\rm e} = 10^6$ $\rm cm^{-3}$. When we take the simple average of the H$\alpha$/H$\beta$, H$\delta$/H$\beta$, H8/H$\beta$, Pa9/H$\beta$, and Pa10/H$\beta$ ratios used in this study, the value changes only slightly from 0.65 at a density of $n_{\rm e} = 10^3$ $\rm cm^{-3}$ to 0.64 at $n_{\rm e} = 10^6$ $\rm cm^{-3}$. Therefore, to facilitate comparison with other studies, we adopted the value at a density of $10^3$, which is commonly used as a standard value. We derived $E(B-V)$ 
from each of the lines listed above,
and subsequently computed an error-weighted average (\autoref{fig:ebv_Center}). When averaging $E(B-V)$, we excluded the very broad component of H$\alpha$ for Center, NE, and SE regions, as including it consistently produced a higher value of $E(B-V)$ from H$\alpha$ than from the other lines. We note that there is a systematic uncertainty of $\sim 5.5-7.5 \:\%$ in extinction-corrected flux that has not been included in the process of reddening correction. 
This uncertainty will affect all lines in a similar manner, and will therefore have a modest impact on derived line ratios.

\section{Results} \label{sec:results}
 In this section, we focus mainly on kinematics and gas properties of the Center region. When deriving gas properties, we use PyNeb version 1.1.17 for electron temperature and density diagnostics.

\subsection{Line identification} \label{ssec:line_identification}
We have detected 57 rest-optical emission lines (\autoref{tab:blue_flux} and \autoref{tab:red_flux}), including auroral lines of four species (\autoref{fig:full_spectrum}). Contrary to previous observations that reported an absence of Balmer lines \citep{vanzella2020}, we detected plentiful Balmer and Paschen lines with S/N of up to $\sim 140$. Interestingly, we detected a very strong permitted \ion{O}{i} \lam8449. This line has been reported in \cite{strom2023} from the stacked spectrum of $z\sim1-3$ galaxies, which they described as unexpected, since metal recombination lines usually are too weak and hardly detected even in local galaxies. Our detection is the first from the spectrum of a single object at such a high redshift. We will discuss this line in Section~\ref{ssec:bowen} and Section~\ref{ssec:bowen_discussion}. In \autoref{tab:blue_flux} and \autoref{tab:red_flux}, we list all emission lines and their flux at Center, normalized by the total flux of H$\beta$. In this paper, we only use the total flux of each line, deferring component-by-component analysis to future work.

\subsection{Kinematics at Center} \label{ssec:kinematics}
\begin{table*}[]
\centering
\caption{Kinematics information for all regions. NE and SE regions only have narrow component since we did one component Gaussian fitting for these regions.}
\begin{tabular}{ccccccc}
\\ \toprule
 & & $z$ & & &FWHM [km s$^{-1}$]& \\
\cmidrule(lr){2-4} \cmidrule(lr){5-7}
       & Narrow  & Broad & Very broad (H$\alpha$) & Narrow & Broad  & Very broad (H$\alpha$) \\ \midrule
\\
\multicolumn{7}{l}{\textbf{(G140H/F100LP)}} \\
Center & 2.36976$\pm$0.00004 & 2.37022$\pm$0.00003 & --- & 37$\pm$26   & 216$\pm$6  & --- \\
NE     & 2.37017$\pm$0.00003 & ---                   & --- & 227$\pm$7 & ---              & --- \\
SE     & 2.37005$\pm$0.00002 & ---                   & --- & 152$\pm$4 & ---              & --- \\
NW     & 2.37070$\pm$0.00002 & 2.36897$\pm$0.00008 & --- & 132$\pm$8   & 376$\pm$8  & --- \\
SW     & 2.37094$\pm$0.00002 & 2.36930$\pm$0.00016 & --- & 126$\pm$7   & 389$\pm$18 & --- \\
\\
\multicolumn{7}{l}{\textbf{(G235H/F170LP)}} \\ 
Center & 2.36990$\pm$0.00004 & 2.37029$\pm$0.00013 & 2.36929$\pm$0.00015   & 110$\pm$22 & 304$\pm$35 & 547$\pm$31 \\
NE & 2.37012$\pm$0.00003     & ---                 & 2.36855$\pm$0.00053   & 239$\pm$8  & ---        & 478$\pm$72 \\
SE & 2.37009$\pm$0.00003     & ---                 & 2.36981$\pm$0.00006   & 126$\pm$9  & ---        & 435$\pm$16 \\
NW & 2.37064$\pm$0.00003     & 2.36903$\pm$0.00009 & ---                   &  93$\pm$16 & 380$\pm$9  & ---              \\
SW & 2.37090$\pm$0.00002     & 2.36918$\pm$0.00013 & ---                   &  92$\pm$11 & 396$\pm$14 & --- \\ 
\\ \bottomrule
\end{tabular}
\label{tab:kinematics}
\end{table*}

As described in Section~\ref{ssec:flux}, we computed kinematic properties from the [\ion{O}{iii}]\lam\lam4960,5008 lines, summarized in \autoref{tab:kinematics}. 
We find a redshift for the narrow component of the [\ion{O}{iii}] doublet of $z_{\text{neb}} = 2.36983 \pm 0.00003$, averaged from the two gratings (see \autoref{tab:kinematics}. This is slightly blue-shifted compared to the value of 2.37025 reported by \cite{mainali2022}, which has been measured from [\ion{O}{iii}]\lam5008. However, \citeauthor{mainali2022} used ground-based spectra, integrating and averaging over a larger area of the sky. IFU velocity maps (Rivera-Thorsen et al. in prep.) show that Godzilla is blue-shifted relative to the systemic redshift.

In the G140H/F100LP grating, the narrow and broad component showed full width at half maximum (FWHM) of $ 37\pm26$ \kms\ and $ 216\pm6$ \kms, respectively. In the G235H/F170LP grating, the FWHMs of the narrow and broad components were $ 110\pm22$ and $ 304\pm35$ \kms. These are values corrected for the instrumental resolution. We suspect that the discrepancy in FWHM between the two gratings is the result of the poorer resolution of the G235H grating, and possibly from inaccuracies in the dispersion calibration files, which are all pre-flight. Although the FWHM is different in the two gratings, the measured fluxes of [\ion{O}{iii}]\lam\lam4960,5008 are consistent within a few percent (\autoref{tab:blue_flux},~\ref{tab:red_flux}). A very broad component observed in H$\alpha$ is blue-shifted by $43\pm14$ and $83\pm14$ \kms\ relative to the narrow and the broad component at the rest-frame ($z=2.36929\pm0.00015$), and has a large FWHM of $547\pm31$ \kms. We found no evidence of an ongoing eruption. First, there is no clear P Cygni profiles in hydrogen and helium lines. Moreover, H$\alpha$ at Center showed a maximum velocity of $\sim 1200$ \kms\ (\autoref{fig:broad_ha_symlog}) which is an order of magnitude smaller than the maximum velocity of H$\alpha$ observed in $\eta$ Car during Great Eruption \citep[$\sim 10^4$ \kms,][]{smith2018}. 

\subsection{Temperature and density diagnostics} \label{ssec:temden}

   \begin{figure}
   \centering
   \includegraphics[width=\columnwidth]{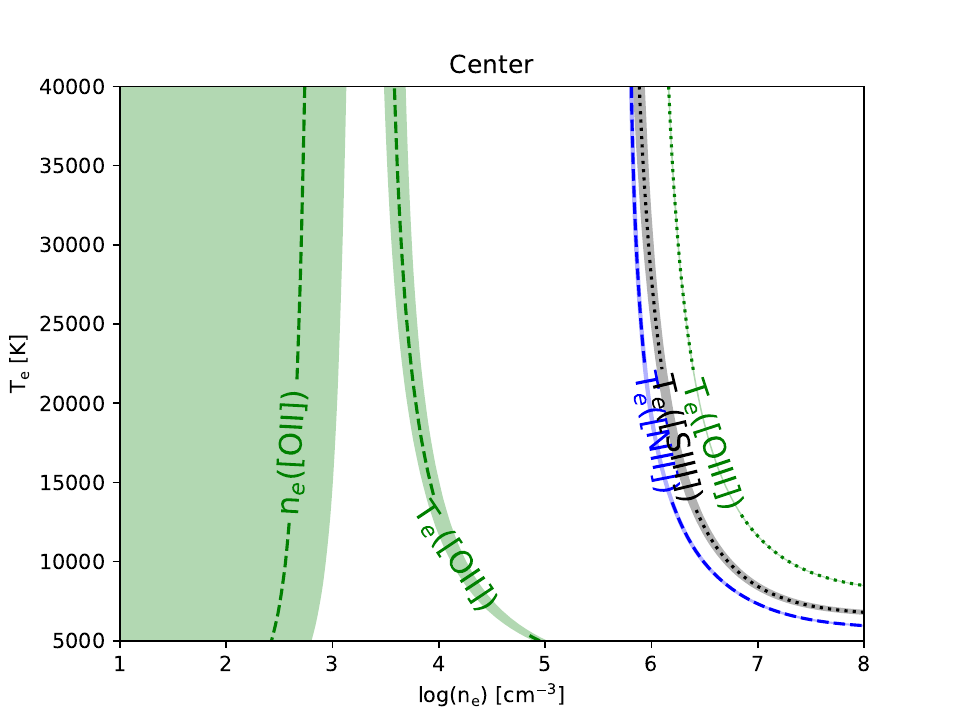}
   \caption{Temperature and density solutions from PyNeb for the spectrum extracted from the Center region in Figure~\ref{fig:five_regions}. Oxygen, sulphur, nitrogen diagnostics are depicted in green, grey, and blue colors, respectively, with shades representing the 1-$\sigma$ confidence regions. We note that the $T_{\rm e} [\ion{O}{ii}]$ diagnostic shown here might not be trustworthy, if indeed the [\ion{O}{ii}]\lam\lam3727,3730 emission has a different region of origin than its auroral counterpart (see Section~\ref{ssec:flux}, Section~\ref{ssec:temden}). Similar figures for the surrounding regions are shown in \autoref{fig:temden_wings}.}
              \label{fig:center_temden}%
    \end{figure}

We have detected auroral lines from four species: [\ion{N}{ii}]\lam5756, [\ion{O}{ii}]\lam\lam7322,7332, [\ion{O}{iii}]\lam4363, and [\ion{S}{iii}]\lam6314. We note that the auroral line of [\ion{O}{iii}]$\lambda\,4363$ is unusually high. 
As density diagnostics, we use [\ion{O}{ii}]\lam\lam3727,3730 and [\ion{S}{ii}]\lam\lam6718,6733, but the S/N of [\ion{S}{ii}]\lam\lam6718,6733 is too low in some regions, including Center. As temperature diagnostics, we use [\ion{O}{ii}]\lam\lam3727,3730/[\ion{O}{ii}]\lam\lam7322,7732, [\ion{N}{ii}]\lam5756/ [\ion{N}{ii}]\lam\lam6550,6585, [\ion{S}{iii}]\lam6314/[\ion{S}{iii}]\lam9071, and [\ion{O}{iii}]\lam4363/ [\ion{O}{iii}]\lam\lam4960,5008.

At Center, PyNeb fails to find a convergent density and temperature solution using the observed diagnostics. \autoref{fig:center_temden} clearly shows that density and temperature diagnostics are probing different gas phases. To explain the observed ratios of the auroral [\ion{O}{iii}]\lam4363, [\ion{S}{iii}]\lam6314, and [\ion{N}{ii}]\lam5756 lines to their nebular counterparts, the gas should have density of $\gtrsim 10^6\:\rm cm^{-3}$. The density range probed by [\ion{O}{ii}]\lam\lam7322,7332 extends down to $\sim10^4 - 10^5 \:\rm cm^{-3}$, but still does not converge with the density solution from [\ion{O}{ii}]\lam\lam3727,3730 of $\lesssim 10^3\:\rm cm^{-3}$. We emphasize that the narrow component of [\ion{O}{ii}]\lam\lam3727,3730 is red-shifted by $140\pm50$ \kms\ in the rest-frame compared to the narrow component of [\ion{O}{iii}]\lam\lam4960,5008. [\ion{O}{ii}]\lam\lam3727,3730 might have an origin different from other emission lines, and may not be suitable as a density diagnostic. Previous observations found that the UV density diagnostic lines \ion{C}{iii}]\lam\lam1907,1909 and \ion{Si}{iii}]\lam\lam1883,1892 suggest a very high density of $n_{\rm e} \gtrsim 10^6 \:\rm cm^{-3}$ \citep{vanzella2020} in Godzilla. One of the main reasons for this discrepancy is that collisionally excited [\ion{O}{ii}] and [\ion{S}{ii}] start to be suppressed at relatively lower density ($n_{\rm crit} \approx 4000-10^4 \: {\rm cm}^{-3}$, \citealt{2011piim.book.....D}) compared to other higher density tracers such as \ion{C}{iii}] \citep{kewley2019}. We discuss this more in Section~\ref{sssec:spatial_analysis}.

\subsection{Bowen fluorescent lines} \label{ssec:bowen}
As reported in Section~\ref{ssec:line_identification}, we detect a strong \ion{O}{i} \lam8449 line at Center, which has flux that is 20.75$\pm$0.15 \% of the H$\beta$ flux, with S/N $\approx 139$. We believe this line has mainly arisen from photo-excitation by accidental resonance (PAR) due to \lyb\ photons, as suggested in \cite{johansson2005} and as reported for \lya-pumped lines in \cite{vanzella2020}. The spectroscopic data disfavors other mechanisms, such as recombination, collisional excitation, or pumping by stellar continuum, considering very weak or absent \ion{O}{i} \lam7776 and \ion{O}{i} \lam7990, and the detection of rest-UV \ion{O}{i} emission lines arising from PAR.

\begin{figure*}
   \centering
   \includegraphics[width=\textwidth]{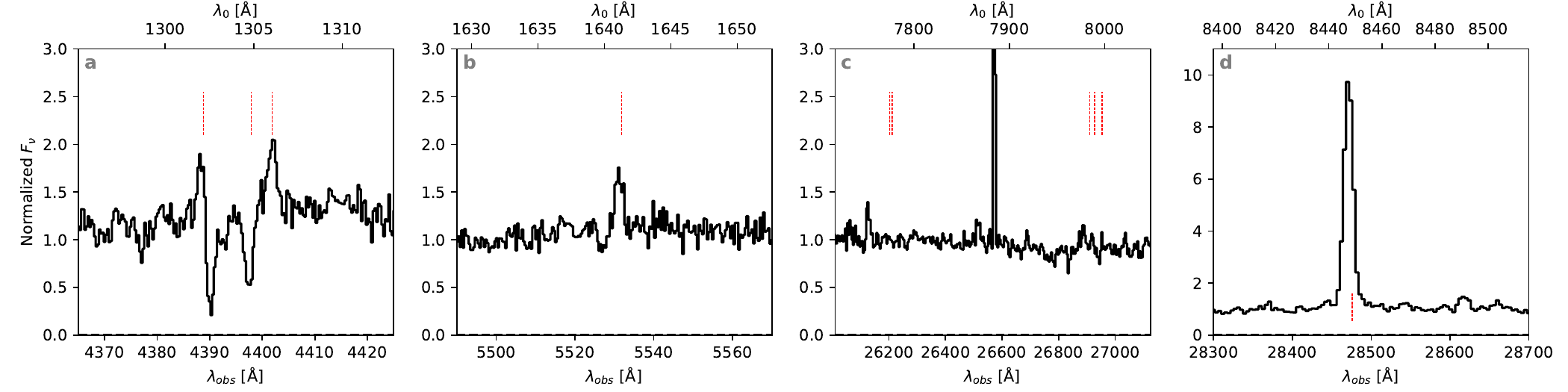}
   \caption{\emph{a,b}: Rest-frame UV spectrum of Godzilla from Magellan/MagE (Rigby et al.\ in prep.), showing the \ion{O}{i} triplet and \ion{Si}{ii} line complex around 1304 \AA\ and \ion{O}{i}] 1641 \AA. All these \ion{O}{i} lines are part of the Ly$\beta$ pumped Bowen fluorescence cascade. \emph{c}: NIRSpec spectrum showing position of the rest-frame optical \ion{O}{i} recombination lines and/or stellar continuum excitation cascade triplet around 7776 \AA\ and line complex at around 7990 \AA, where we would expect to see line emission if the observed \ion{O}{i} cascade were due to recombination or stellar continuum pumping. \emph{d}: \lyb-pumped, Bowen fluorescent \ion{O}{i} 8449 \AA.}
              \label{fig:OI_MagE}%
    \end{figure*}

\autoref{fig:OI_MagE} shows various \ion{O}{i} lines in the rest-frame UV (\emph{a}, \emph{b}) and optical (\emph{c}, \emph{d}) wavelength. Panel \emph{c} shows an absence of the \ion{O}{i} \lam7776 triplet and \ion{O}{i} \lam7990 complex. Measuring the ratio between \ion{O}{i} $\lambda$7776/$\lambda$8449 yields a $3 \sigma$ upper limit of 0.009. 
This is inconsistent with the first two scenarios, since we expect \lam 7776/\lam8449 $\approx 1.7$ and \lam7776/\lam8449 $\approx 0.3$ for the case of recombination and collisional excitation, respectively \citep{grandi1980, haisch1977}. The line ratio \ion{O}{i} \lam7990/\lam8449 is expected to have a value of \lam7990/\lam8449 $\approx 0.052$ from the continuum fluorescence cascade calculation by \cite{grandi1980} (assuming case B recombination and $T_{\rm e} = 10^4$ K). This is is inconsistent with the observed value $0.013\pm0.005$ in Godzilla, by more than $7\sigma$, significantly weakening the stellar continuum excitation scenario. Assuming case A recombination, these ratios can become as low as 0.0030, but case A does not seem to be a proper assumption as we already know that there exists a dense gas component. \ion{O}{i} \lam7256 is another line emitted from the same stellar continuum excitation cascade, but it falls in the detector gap.

Moreover, we point out that the UV counterparts of \lyb-pumped lines, the \ion{O}{i} 1302.2, 1304.9, 1306.0 triplet and \ion{O}{i}]\lam1641.31 \citep[see their Figure 1]{shore2010} emission lines, are detected in the MagE spectrum, although 1304.9 is not obvious due to the effect of Si II absorption (\autoref{fig:OI_MagE}-\emph{a}, \emph{b}). The observation of these other \lyb-pumped lines of the same cascade, as well as the absence of recombination lines and continuum-pumped lines, supports that the permitted \ion{O}{i} \lam8449 is a \lyb-pumped line. Previously, emission near 1640 \AA\ was identified as \ion{He}{ii} \lam1640.42 by \cite{vanzella2020} and suggested as supporting evidence for a transient scenario. On the other hand, \cite{diego2022} argued that it is \ion{O}{i}]\lam1641 based on its line center. Taking into account that we have detected other \lyb-pumped \ion{O}{i} lines while not detecting any \ion{He}{ii} emission lines, we conclude that this emission is from \ion{O}{i}.

We also report the detection of \lya-pumped \ion{Fe}{ii} \lam8490 with S/N of 8. Interestingly, the other \lya-pumped line near this line, \ion{Fe}{ii} \lam8453, has not been detected. Discussion of the Bowen fluorescent lines will be continued in Section ~\ref{ssec:bowen_discussion}.

\section{Discussion} \label{sec:discussion}
\begin{figure}
   \centering
\includegraphics[width=\columnwidth]{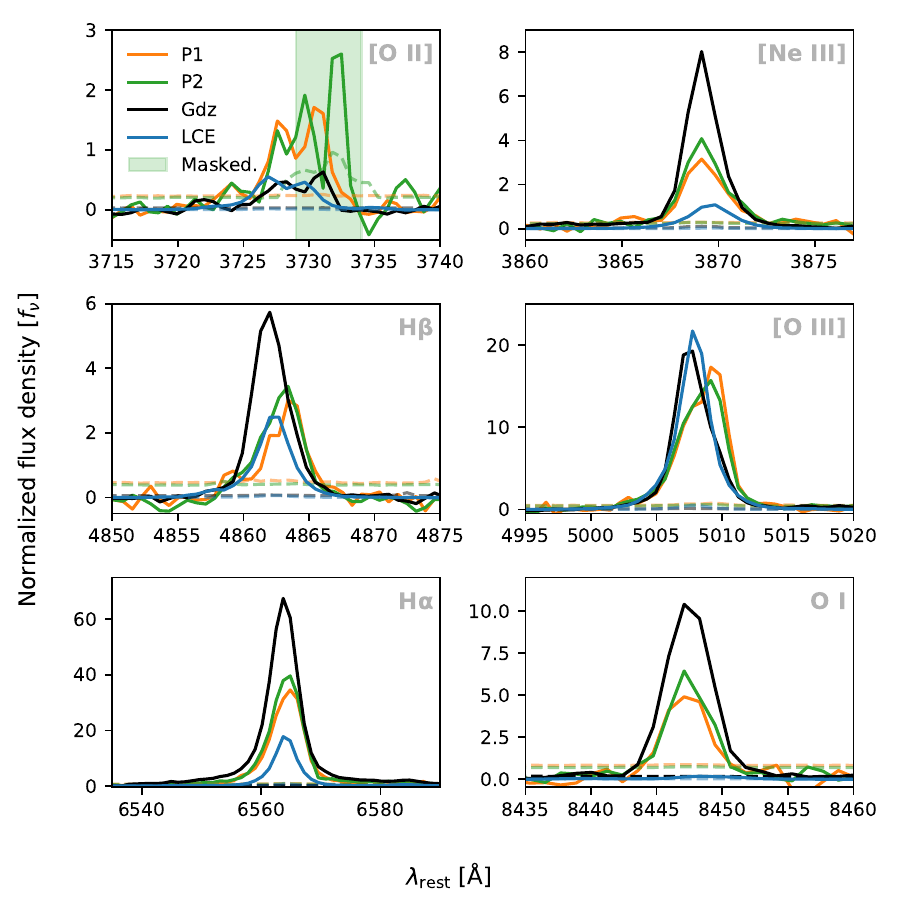}
   \caption{Selected, relative line fluxes from Godzilla, the two \emph{P}-knots, and for comparison the Lyman Continuum emitter (LCE) knot, highlighting kinematics and relative line strengths. The green shaded interval in the [\ion{O}{ii}] panel indicates where P2 suffers from contamination from noisy pixels. The spectra were first normalized by their median value, the continuum then subtracted, and the resulting lines all normalized by the integral of [\ion{O}{iii}] 5008 of each spectrum.}     \label{fig:p_knots_lines}%
    \end{figure}

\begin{figure}
   \centering
\includegraphics[width=\columnwidth]{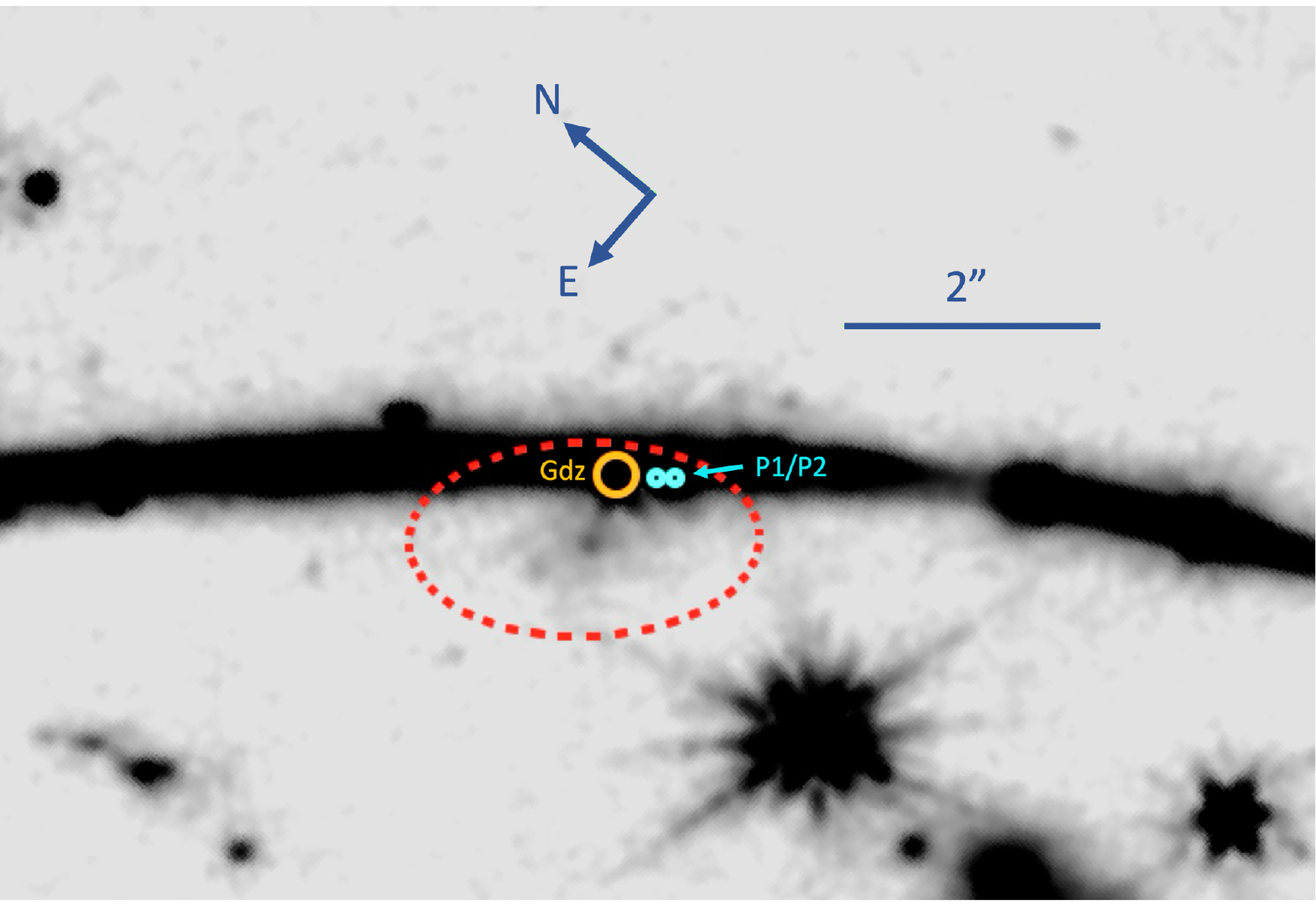}
   \caption{Combined NIRCam F115W+F150W image, smoothed by a FWHM = \ang{;;0.1} Gaussian to bring out low surface brightness details. The dashed ellipsoid marks a low surface brightness galaxy close to Godzilla (larger circle) and the P knots (smaller circles).}     \label{fig:perturber}%
    \end{figure}

\subsection{Revisiting the transient scenario and the extreme magnification scenario} \label{ssec:lensing}

\subsubsection{Transient scenario}
The NIRCam data weakens the SN scenario suggested by \cite{vanzella2020}. \autoref{fig:nircam_image} shows the April 2023 NIRCam observation, and Godzilla still maintains similar brightness to that of the Sunburst LCE (clump 1, marked with red squares). It expands the observed duration of Godzilla from previously 2 years to 5 years. While \cite{diego2022} included the original discovery ESO New Technology Telescope (NTT) observations in 2014 \citep{dahle2016}, we exclude this as there Godzilla is not clearly resolved. Counting from the February 2018 HST observation, Godzilla has maintained its approximate luminosity for $\sim$1.5 years in the rest-frame.

\cite{vanzella2020} argued that Godzilla (``Tr'' in their terminology) is a Type IIn SN, a type of SN with narrow lines in their spectra. These narrow lines are believed to arise when a massive star explodes into a dense CSM. Although Type IIn SNe can be observed more than a few decades after explosion \citep{immler2005,milisavljevic2012}, it does not mean that it maintains high luminosity for that long time. One sub-type of Type IIn SNe, IIn-P \citep{mauerhan2013}, is a type that shows  a lasting, luminous plateau phase.
A recent theoretical work however predicted the duration of this plateau phase to around 100 days at most \citep{khatami2023}. Thus, a luminous plateau lasting for at least 1.5 years observed in Godzilla is highly unlikely to happen in any kind of SN known so far.

Godzilla is also unlikely to be some other kind of rare transient that shows an approximately flat light curve for 1.5 years. As summarized in Section~\ref{sec:introduction}, such a scenario contradicts the maximum time delay predicted from the lens model by \cite{sharon2022}, as a sudden increase in luminosity should have been observed in counter images in other arcs.

\subsubsection{Extreme magnification scenario}
Godzilla and a pair of clumps next to it (the ``P knots'' following \citealt{diego2022}) are believed to be nearby in the source plane \citep{diego2022}. An extreme magnification can happen if Godzilla lies on the critical curve, as suggested in \cite{diego2022}. In \autoref{fig:p_knots_lines}, we compared emission line profiles normalized by the flux of [\ion{O}{iii}]\lam5008 at each spectrum for the P knots, Godzilla and the Sunburst LCE. The left and right clumps in the P knots, P1 and P2, show nearly identical line profile traits such as redshift, line ratios and line width in various emission lines, and are clearly distinguished from Godzilla. It suggests that the P knots are likely to be the mirrored image of the same object different from Godzilla, and strengthens the hypothesis of a perturbing mass in the foreground generating a critical curve crossing the top of Godzilla and between the P knots \citep[see Figure 6]{diego2022}. 

The plausibility of this lensing scenario is strengthened by the NIRCam detection of a low surface brightness galaxy centered only \ang{;;0.5} from Godzilla. As shown in \autoref{fig:perturber}, this galaxy has a clumpy structure and may be responsible for creating the small-scale perturbations needed to shift the critical curve to the locations suggested by  \cite{diego2022}.



\subsection{Magnification factor derived from counterimages} \label{ssec:magnification}

\begin{figure}
   \centering
\includegraphics[width=\columnwidth]{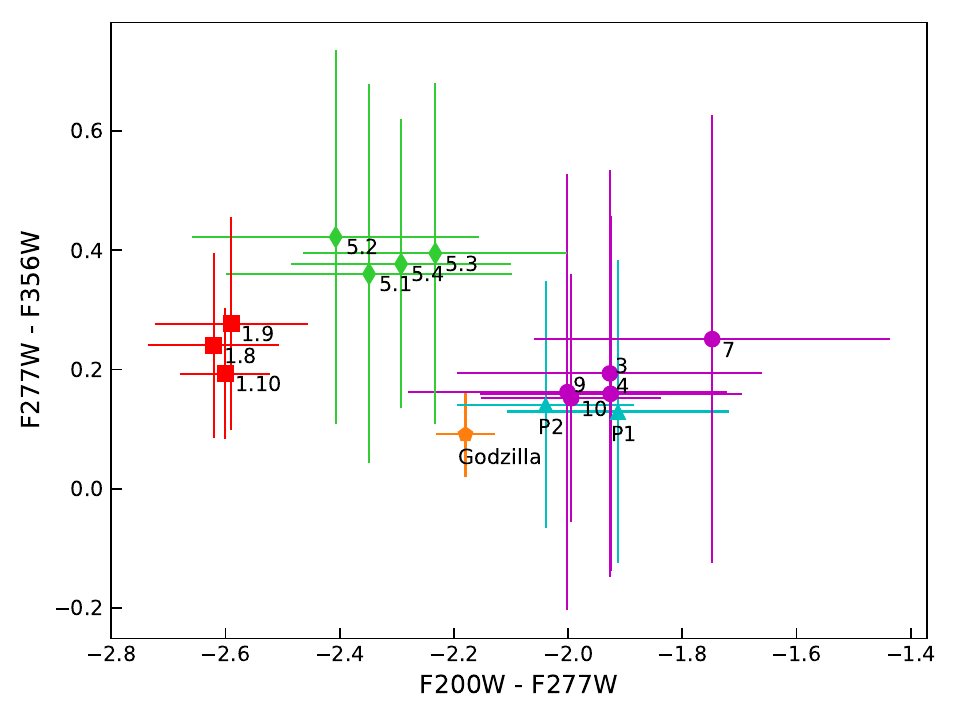}
   \caption{A color-color diagram using the NIRCam F200W, F277W, and F356W filters. Error bars are shown for each point. In the color-color space, Godzilla (orange pentagon) is positioned closer to the images of clump 4 (purple circles, `4.' in `4.X' has been omitted) and P knots (cyan triangles). Its location is clearly distinguished from that of the images of clump 1 (red squares) and clump 5 (green thin diamonds).}\label{fig:color_color}
    \end{figure}

\begin{figure*}
   \centering
\includegraphics[width=\textwidth]{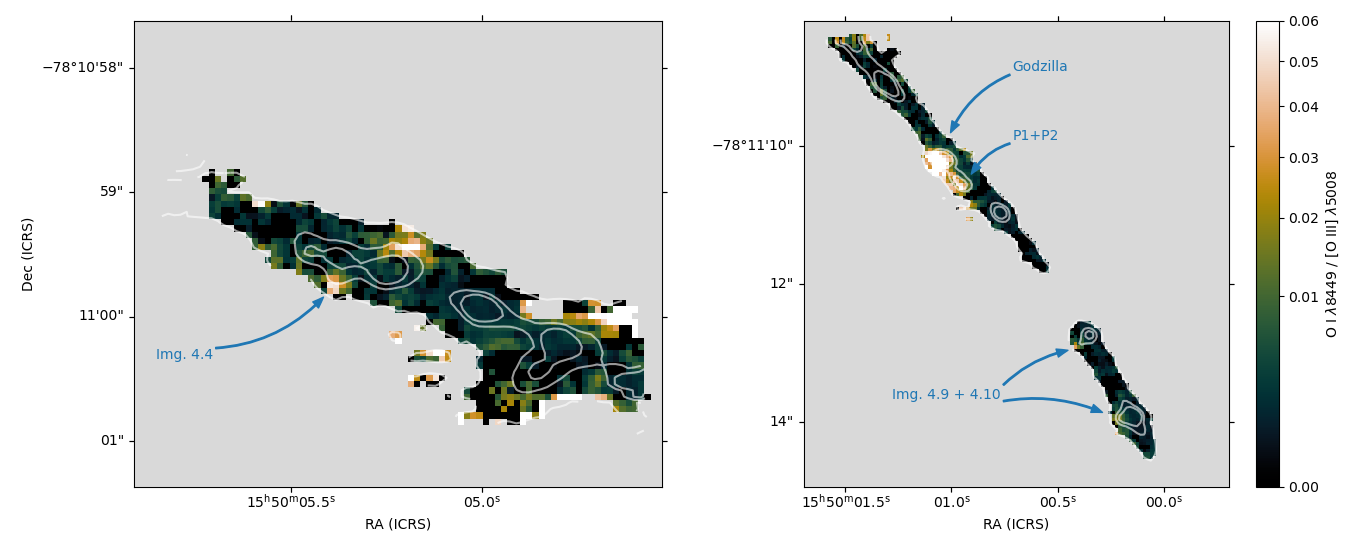}
   \caption{\ion{O}{i} \lam8449/[\ion{O}{iii}]\lam5008 map for Pointing 1 (left) and Pointings 2+3 (right), with the stellar continuum overlaid as contours. The \ion{O}{i} \lam8449/[\ion{O}{iii}]\lam5008 ratio is near zero in the images of clump 1 and is highest in the images of clump 4, Godzilla, and the P knots. Refer to $\autoref{fig:O1_Hb_map}$ in the appendix to see the map normalized by H$\beta$.}\label{fig:O1_O3_map}
    \end{figure*}

   \begin{figure}
   \centering
\includegraphics[width=\columnwidth]{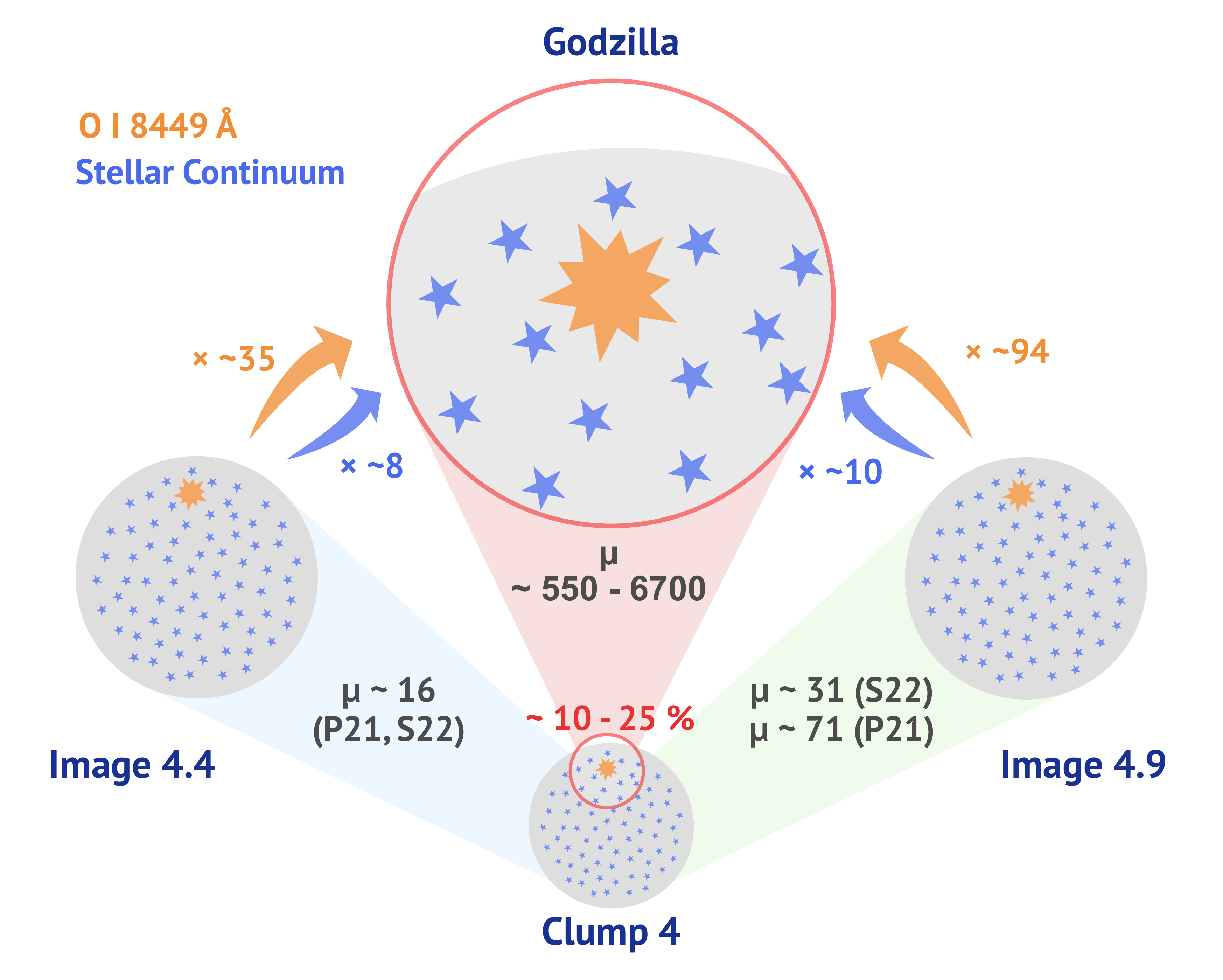}
   \caption{An illustration explaining the relationships between clump 4, its two counterimages (images 4.4 and 4.9), and Godzilla. The difference between the \ion{O}{i} \lam8449 flux ratio and the stellar continuum flux ratio indicates that Godzilla contains only a portion of the stars from clump 4, representing between $\sim10$ \% and $\sim 25$ \% of the total stellar light, depending on the model and the image used for comparison. The magnification factor also varies significantly, ranging from $\sim 560$ to $\sim 6700$ (excluding errors).}\label{fig:conceptual_figure}
    \end{figure}

\begin{table*}[]
\centering
\caption{
Flux ratio compared to candidate counterimages and the derived magnification factor. In the table, the top row shows the comparison with image 4.4, and the bottom row shows the comparison with image 4.9. The continuum flux is calculated as the average of the flux ratios from 6 NIRCam filters (F115W, F150W, F200W, F277W, F356W, F444W), weighted with the inverse square of the error. The O I flux was calculated from each clump’s one-dimensional spectrum. The magnification factor was determined by comparing the values for images 4.4 and 4.9 given in \cite{pignataro2021} and \cite{sharon2022} and denoted as P21 and S22, respectively.}
\label{tab:conterimages_magnification}
\begin{tabular}{lllll}
\\ \toprule
           & $f/f_{4.4}$ (Continuum) 
           & $f/f_{4.4}$ (\ion{O}{i} 8449$\AA$) & $\mu$ (P21) &  $\mu$ (S22)       \\ \midrule
Godzilla   & $7.6\pm0.38$ & $35\pm5.1$ & $570^{+100}_{-100}$ & $550^{+92}_{-120}$\\
P1         & $1.2\pm0.077$ & $1.2\pm0.20$ & $20^{+3.8}_{-3.9}$ & $19^{+3.5}_{-4.5}$ \\
P2         & $1.7\pm0.099$ & $1.6\pm0.25$ & $26^{+4.9}_{-5.0}$ & $25^{+4.4}_{-5.9}$ \\
\\ \midrule
           & $f/f_{4.9}$ (Continuum) 
           & $f/f_{4.9}$ (\ion{O}{i} 8449$\AA$) & $\mu$ (P21) &  $\mu$ (S22)       \\ \midrule
Godzilla   & $9.5\pm0.55$ & $94\pm22$ & $6700^{+1700}_{-1700}$ & $2900^{+1100}_{-740}$\\
P1         & $1.5\pm0.10$ & $3.3\pm0.8$ & $230^{+61}_{-60}$ & $100^{+40}_{-26}$ \\
P2         & $2.1\pm0.14$ & $4.3\pm1.0$ & $310^{+80}_{-78}$ & $130^{+50}_{-34}$ \\ \bottomrule
\end{tabular}
\end{table*}

\subsubsection{Counterimages of Godzilla}
\cite{diego2022} suggested images 5.1 - 5.4 (marked with green circles in Figure 2) and 4.7 as candidate counterimages of Godzilla based on that they are located between images of clumps 1 and 2. On the other hand, the lens model from \cite{sharon2022} predicts that Godzilla and the P knots together make up the highly resolved 8th counterimage of Clump 4 (depicted with purple circle labeled ``4.8'' in \autoref{fig:nircam_image}). 

To identify possible counterimages of Godzilla, which should have a similar SED as Godzilla, we examined the SEDs for Godzilla, the P knots, and the images of clumps 1, 4, and 5. Using \texttt{Photutils} \citep{larrybradley2024}, we extracted flux from 0.06 arcsec radius circular apertures, using the images taken with 6 NIRCam filters (F115W, F150W, F200W, F277W, F356W, F444W). A color-color diagram was then created by combining the filters that displayed the most distinct differences in SED shape (F200W-F277W versus F277W-F356W), as shown in \autoref{fig:color_color}. The stellar continuum properties of Godzilla appear to be closer to clump 4 than to clump 5. The P knots also occupy a position on the color-color diagram that is indistinguishable from clump 4.

The unusually bright \ion{O}{i} \lam8449 in Godzilla provides an additional method for investigating its counterimages. \autoref{fig:O1_O3_map} shows the \ion{O}{i} \lam8449/[\ion{O}{iii}]\lam5008 map for the north arc and northwest arc, revealing that strong \ion{O}{i} \lam8449 emission is a unique feature of only Godzilla, P knots and clump 4, aside from the bright areas corresponding to image 8.4 in the pointing 1. Of the clump 5 images, only image 5.4, located next to image 1.4, falls within the NIRSpec pointing area and does not show strong \ion{O}{i} \lam8449. To test whether this is due to suppressed [\ion{O}{iii}]\lam5008 caused by high density, we have normalized \ion{O}{i} \lam8449 with H$\beta$ (\autoref{fig:O1_Hb_map}). We observe an O I excess in the same regions as seen in the [\ion{O}{iii}]\lam5008 normalization map, and image 5.4 still shows no elevation. Thus, the elevation in \ion{O}{i} \lam8449 is not the result of suppressed [\ion{O}{iii}]\lam5008 in the high-density region. We believe that the [\ion{O}{iii}]\lam5008 emission observed in the Godzilla region originates from high-density gas with $n_{\rm e} \gtrsim 10^6 \:\rm cm^{-3}$. As shown in \autoref{fig:center_temden}, [\ion{O}{iii}]\lam5008 probes regions with $n_{\rm e} \gtrsim 10^6 \:\rm cm^{-3}$, regardless of temperature. While the \ion{O}{i}/H$\beta$ ratio is also influenced by oxygen abundance, the \ion{O}{i} \lam8449/[\ion{O}{iii}]\lam5008 map, which is independent of metallicity, shows that the elevated \ion{O}{i} \lam8449 is not due to high oxygen abundance. \ion{O}{i} \lam8449 exhibits an elevation due to a unique \lyb-pumping mechanism, which helps constrain the counterimage of Godzilla.

Through photometric color comparison (\autoref{fig:color_color}) and the \ion{O}{i} \lam8449 maps (\autoref{fig:O1_O3_map}, \autoref{fig:O1_Hb_map}), we concluded that images of clump 4 and P knots are counterimages containing Godzilla, as they share a similar stellar continuum color and characteristic nebular emission.


\subsubsection{Discrepancy between the stellar continuum and O I flux ratios in counterimages}\label{sssec:magnification}

The magnification factor of the candidate counterimages is known from the gravitational lens model; therefore, by comparing the flux of the candidate counterimages with that of Godzilla, we can determine Godzilla’s magnification factor. However, simply comparing the stellar continuum brightness is not sufficient. \cite{sharon2022} considers Godzilla and the P knots as an enlarged version of clump 4, which makes direct flux comparison difficult if Godzilla is indeed a part of clump 4. Since \ion{O}{i} \lam8449 is a very unique line, if we assume that the only source of \ion{O}{i} \lam8449 observed in clump 4 is inside Godzilla and that there are no other major \ion{O}{i} \lam8449 sources, comparing the brightness of \ion{O}{i} \lam8449 between the clump 4 images and Godzilla would provide a more accurate comparison.

We extracted one-dimensional spectra from Godzilla, the two P knots (P1, P2), and images 4.4 and 4.9. We did not include image 4.10, as its close proximity to the bright image 1.10 might cause contamination. To measure the total \ion{O}{i} \lam8449 flux for Godzilla, we re-extracted its spectrum over a $7\times5$ pixel area that includes the five regions (Center, NE, SE, NW, SW) shown in \autoref{fig:five_regions}. The continua have been removed following the method described in Section~\ref{ssec:extraction}, then we summed the flux of the \ion{O}{i} \lam8449 line within the extraction area. We also compared the flux of the stellar continuum using photometry measurement employed for SED and color comparison in the previous section. The results are shown in \autoref{tab:conterimages_magnification}.

Godzilla exhibits a large discrepancy between its continuum flux ratio and its \ion{O}{i} \lam8449 flux ratio. Compared to image 4.4, Godzilla’s continuum flux is 7.6 times brighter, while its \ion{O}{i} \lam8449 flux is much brighter than that of image 4.4 with the ratio of 35. Assuming that Godzilla is the sole \ion{O}{i} \lam8449 source, a flux ratio of 35 would be a more likely measure of the true magnification ratio between the two images. Under the assumption of a homogeneous magnification across the Godzilla region, the fact that the stellar continuum is only 7.6 times brighter suggests that the Godzilla region contains $\sim22 \%$ ($7.6/35\times100$) of the stars compared to image 4.4. Meanwhile, P1 and P2 do not show significant differences between their continuum flux ratios and \ion{O}{i} \lam8449 flux ratios when they were compared to image 4.4. From this, we can infer that P1 and P2 represent the entirety of image 4.4, rather than just a part of it. The magnification factor for image 4.4 has derived to be $16.2^{+1.7}_{-1.8}$ from \cite{pignataro2021} and $15.7^{+1.3}_{-2.7}$ from \cite{sharon2022}. Combining these magnification factors with the \ion{O}{i} flux ratio, the magnification factors for Godzilla, P1, and P2 are inferred to be $\mu \approx 560$, $\mu \approx 20$, and $\mu \approx 25$, respectively.

Comparing to image 4.9 presents a slightly different picture. The difference between the stellar continuum and \ion{O}{i} \lam8449 flux in Godzilla is even more pronounced in this comparison, where the continuum flux ratio is $\sim 9.5$ and the \ion{O}{i} \lam8449 flux ratio soars to $\sim 94$. Interpreting this with the same logic as before suggests that Godzilla contains only $\sim 10$ \% of the stars present in clump 4. In the meanwhile, P1 and P2 show discrepancy of only factor of 2. The magnification factors for image 4.9 vary significantly between the two models presented. \cite{pignataro2021} reported the magnification factor for image 4.9 to be $71.1^{+7.6}_{-6.4}$ while \cite{sharon2022} reported it as $31.1^{+8.9}_{-3.1}$, noting the magnifications of images 1.9, 1.10, 4.9, and 4.10 should not be taken at face value. The magnification factors for Godzilla are then inferred to be $\mu \approx 6700$ and $\mu \approx 2900$ based on the models from \cite{pignataro2021} and \cite{sharon2022}. This is all summarized in \autoref{fig:conceptual_figure}, which clearly displays the variations in Godzilla's magnification factor and the percentage of stellar components based on different images and models.

Such widely varying values indicate that the current lens model has not reached a consensus. In image 4.4, predictions from the two models were similar and the critical line in that area was well constrained. However, it is still unclear if the magnification factor for image 4.4 is indeed more accurate than that for image 4.9. The stellar continuum flux in image 4.4 is $1.2\pm0.09$ times brighter than that of 4.9, yet the lens models suggest magnification factor for image 4.4 to be $\sim 1/2-1/5$ smaller than that of 4.9, indicating an internal inconsistency. Regarding the differences between models, despite symmetry between images 9 and 10, a model by \cite{sharon2022} does not place the critical line between them. This results in a lower magnification factor than \cite{pignataro2021}, which used a perturber (galaxy 1298 in that study) to bend the critical line. Recently, the updated lens model by \cite{solhaug2024} proposed a new perturber location (Perturber I therein) that shifts the critical line to run between images 9 and 10. This work reports a magnification factor $\sim270$ for image 4.9 (E. Solhaug, private communication), which leads to $\mu\approx25000$ for Godzilla.
This is an extreme number compared to $\mu\approx600 - 7000$ derived by \cite{diego2022} or $\mu\approx190 - 5000$ by \cite{pascale2024} based on the same counterimage comparison methods. It should also be noted that results from \cite{diego2022} are based on the assumption that clump 5 is the counterimage of Godzilla. While Pascale used clump 4 instead, both \cite{diego2022} and \cite{pascale2024} lacked observational data on \ion{O}{i} \lam8449 and thus compared only the photometric brightness of Godzilla and its counterimage. They assumed that the total brightness of the counterimage corresponds to the total brightness of the Godzilla region and did not consider the possibility that Godzilla corresponds to only a part of the clump. 

These two studies also aimed to estimate the maximum magnification of Godzilla. \cite{diego2022} suggested that while magnifications up to $10^5$ due to small-scale perturbers such as stars are theoretically possible, there are two key limitations: (1) such extreme magnifications cannot be sustained over long periods due to the motion between the star and the caustic, and (2) microlenses within the cluster can reduce the maximum achievable magnification, especially when the magnification is very high and caustics are overlapping ($\tau_{\text{eff}} \gg 1$). Based on these factors, the highest sustained magnification was estimated to be a few times $10^4$. On the other hand, \cite{pascale2024}, based on microlensing simulations and the fact that Godzilla shows flux variations of less than 3\%, concluded that the magnification factor of Godzilla is unlikely to exceed 2000. However, their approach relies on the value $\log(\mu M_\star/M_\odot) = 9.3$. Adopting their magnification factor, it means that the stellar mass of the Godzilla should be comparable to that of the LCE cluster. Comparing the rest-frame optical images of clump 4 and the LCE cluster with similar magnification (e.g. image 1.4 and 4.4 where $\mu=15.3^{+7.7}_{-6}$ and $\mu=15.7^{+1.3}_{-2.7}$, respectively, according to \cite{sharon2022}), their relative brightnesses are not consistent with being of similar mass, and it makes their maximum magnification estimation less convincing. 


These contradictory results from different studies illustrate how the estimated magnification factor of Godzilla can vary significantly depending on the chosen lens model. Given the current lens models, we cannot rule out the scenario in which Godzilla is magnified by a very high factor of $\sim 10^4$, which represents the highest sustainable magnification.

\subsection{\ion{O}{i} \lam8449 emitting source}
\subsubsection{Spatial analysis} \label{sssec:spatial_analysis}
\begin{table}[]
\centering
\caption{Computed color excess ($E(B-V)$) for the 5 regions.
\label{tab:dust_factors}}
\begin{tabular}{ll}
\\ \toprule
           & $E(B-V)$          \\ \midrule
Center     & 0.45$\pm$0.01 \\
NE         & 0.78$\pm$0.04 \\
SE         & 0.77$\pm$0.03 \\
NW         & 0.24$\pm$0.02 \\
SW         & 0.25$\pm$0.03 \\ \bottomrule
\end{tabular}

\end{table}

Our main interest lies not in the components contributing to Godzilla's ordinary stellar continuum, but in identifying the source responsible for the \ion{O}{i} \lam8449 emission and other unique characteristics of Godzilla. Spatially analyzing the surrounding region around the brightest Center may help understand this source. As seen in \autoref{tab:kinematics}, the broadest component at Center, NE, and SE regions show a high value of $\mathrm{FWHM}\sim435 - 547$ \kms. It is larger than the expansion velocity of most local LBVs (a few to 100 $\rm km\,s^{-1}$), and similar to that of $\eta$ Car \citep{weis2020}. While NW and SW regions do not have the very broad component in H$\alpha$, these regions show strong outflow features in many lines. In the G140H/F100L grating, NW and SW regions show broad components blue-shifted compared to the narrow component by $143\pm8$ \kms\ and $153\pm12$ \kms\ in the rest-frame, respectively.

One distinguishing feature of Godzilla is that it is very dusty. The Sunburst Arc is in general not so dusty \citep{mainali2022}. The dust cover in this region seems extremely uneven; in the spatial dust map of the whole Sunburst Arc, Godzilla and counterimages of clump 4 stand out with a significantly higher $E(B-V)$ than typical values for the galaxy (Rivera-Thorsen et al. in prep.). As shown in \autoref{tab:dust_factors}, the $E(B-V)$ reaches $\sim 0.24 - 0.78$ , depending on the region. The NW and SW regions, where strong outflow features are observed (\autoref{fig:five_regions}), are less dusty. Interestingly, \cite{mainali2022} find that their MagE spectrum containing Godzilla have a slightly lower $E(B-V)$ compared to other regions based on the H$\alpha$/H$\beta$ Balmer decrement (0.02 for the M-3 pointing containing Godzilla vs. 0.04--0.15 for the other pointings). The $E(B-V) \approx 0.02$ for the M-3 pointing, which includes Godzilla, is significantly lower than the value derived in this study. This discrepancy is likely due to aperture dilution. The Balmer emission line map shows strong emission along the core of the arc, with a significantly lower H$\alpha$/H$\beta$ value than in Godzilla (Rivera-Thorsen et al. in prep.). Given that the M-3 slit includes the central spine of the arc adjacent to Godzilla, we anticipate a substantial dilution effect, lowering the calculated E(B-V) value. Moreover, unlike the NIRSpec spectra, MagE spectra are subject to air blurring as they are ground-based, which would further enhance the aperture dilution effect.

Assuming $R_V=4.05$, we obtain $A_V=1.8$ for the Center, a value strikingly close to $A_V\approx2.0$ derived for the Weigelt blobs \citep{davidson1995,hamann1999}. Despite significant dust reddening, strong UV emission lines are observed in the Weigelt blobs, as they are for Godzilla. This may suggest that the dust is mixed with the nebular gas, allowing emission from the outer layers of the gas to escape with minimal attenuation, rather than being absorbed by a dust screen external to the nebular gas.

Given the amount of dust Fe is typically expected to be depleted, yet we detect abundant Fe lines. This phenomenon is also observed in $\eta$ Car, and \cite{smith2007} suggested that Fe-bearing grains can be selectively destroyed while other dust grains remain intact; a similar process might be occurring in this system.

The surrounding regions are also variant in their gas temperature and density. Looking at the temperature and density diagnostics diagram in the surrounding regions (\autoref{fig:temden_wings}), we can see that the density is higher ($n_{\rm e} \gtrsim 10^3 \:\rm cm^{-3}$) in the SE and SW regions compared to the NE and NW regions ($n_{\rm e} < 10^3 \:\rm cm^{-3}$). The cross section of the density diagnostics and low temperature diagnostic line ([\ion{O}{ii}]\lam\lam7322,7332) is also more biased to the higher density in the SE and SW regions.

Spatially variant kinematics, dust properties, and gas temperature and density clearly show that the central bright source is surrounded by multi-phase, inhomogeneous gas. Depending on the magnification, if it is closer to 7000, this may imply a circumstellar medium and dust ejected through stellar winds or previous eruptions. If the magnification is closer to 600 and Godzilla displays a broader region, this could reveal spatial variation within a larger nebular region. 

\subsubsection{\ion{O}{i} \lam8449 source candidates} \label{sssec:excluded_candidates}
   \begin{figure}
   \centering
\includegraphics[width=\columnwidth]{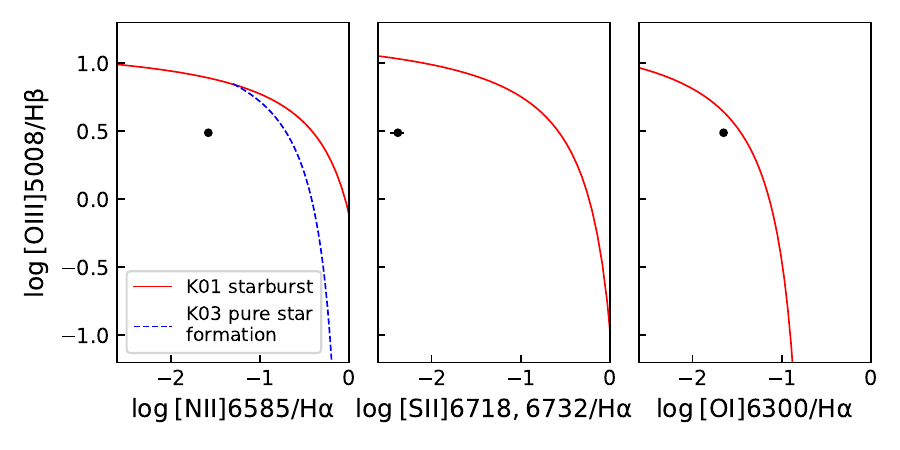}
   \caption{N2, S2, O1 BPT diagram of Center (black dot with error bars). Theoretical maximum starburst \citep{kewley2001} and empirical star formation \citep{kauffmann2003} lines are shown in red solid and blue dashed lines respectively.}
              \label{fig:bpt}%
    \end{figure}

Astronomical objects that emit \ion{O}{i} \lam8449 through the \lyb-pumping mechanism are limited. We exclude \textbf{reflection nebulae and \ion{H}{ii} Regions} since The \ion{O}{i} \lam8449 line observed there is typically produced by stellar continuum pumping or recombination, not by \lyb-pumping. For example, \cite{grandi1975} demonstrated that starlight continuum fluorescence is the preferred excitation mechanism for the \ion{O}{i} line in the Orion Nebula. We also exclude \textbf{supernova remnants (SNRs)}. Certain supernova remnants, especially those interacting with surrounding interstellar material, can exhibit \ion{O}{i} \lam8449 line emissions. However, \ion{O}{i} \lam8449 emission through \lyb-pumping is rare in SNRs because they typically lack the necessary combination of dense neutral oxygen and intense UV radiation near the remnant. For example, from the ratio of \ion{O}{i} \lam7776 and \ion{O}{i} \lam8449, \cite{winkler1985} and \cite{itoh1986} have shown that \ion{O}{i} \lam8449 observed in SNR Puppis A and Cassiopeia A mainly results from recombination. \textbf{Very Massive Stars (VMS) and Wolf-Rayet (WR) stars} are also easily excluded since they are not known for strong Bowen fluorescent lines. Moreover, we detect neither nebular nor broad stellar \ion{He}{ii} and there is no sign of broad stellar \ion{C}{iv} emission lines at 5808 \AA\, which effectively excludes a VMS and WR stars. We also exclude \textbf{classical Be stars}. Classical Be stars are B-type stars of luminosity classes V-III that exhibit prominent Balmer emission lines \citep{jaschek1981}. Be stars are the rapidly rotating stars and we can observe emission lines from the disks or ring-like envelopes surrounding the low-latitude regions \citep{kogure2007}. Although \cite{mathew2012b, mathew2012} have argued that \lyb-pumping plays an important role in the excitation of \ion{O}{i} \lam8449 lines observed in classical Be stars, Figure 4 of \cite{mathew2012b} shows that there is still a significant contribution from collisional excitation, unlike in Godzilla. \textbf{B[e] stars}, a subgroup of peculiar Be stars that exhibit IR excess and forbidden lines \citep{allen1976}, are also excluded due to the lack of prominent \lyb-pumped \ion{O}{i} \lam8449 lines. For example, the spectrum of B[e] star HD 45677 shows \ion{O}{i} $\lambda7776/\lambda8449\approx0.3$ \citep[Fig. 4]{dewinter1997} which can be explained by collisional excitation.

Following is a list of objects that could emit strong \ion{O}{i} \lam8449 mainly through \lyb-pumping:

\begin{itemize}
\item \textbf{Broad-line regions (BLRs) of  active galactic nuclei (AGNs) and quasars:} High UV fluxes from the central black hole can ionize the surrounding gas and produce Bowen fluorescence lines. After analyzing the spectra of 16 Seyfert galaxies, \cite{grandi1980} concluded that 13 of them exhibited broad \ion{O}{i} \lam8449 emission, with \lyb-pumping identified as the excitation mechanism. Godzilla does not display line widths approaching those typically found in BLRs of several thousand km/s; and does not show high-excitation emission lines such as \ion{He}{ii}, \ion{N}{v}, or \ion{O}{vi} which are typical in BLRs of AGNs. In \autoref{fig:bpt}, we show the position of Center of Godzilla in the N2, S2, and O1 BPT diagrams; it clearly falls within the stellar region, without any evidence for high-energy excitation or strong shocks, ruling out the black hole scenario discussed in \cite{diego2022}.

\item \textbf{Planetary nebulae:}
In dense regions close to the central star of planetary nebulae, \lyb\ photons can excite neutral oxygen efficiently. \lyb-pumped \ion{O}{i} \lam8449 emission has been observed in several compact, high-density planetary nebulae, such as NGC 7027 \citep{rudy1992}, IC 5117 \citep{rudy2001}, and IC 4997 \citep{rudy1989, feibelman1992}. However, due to the hot central star, a highly ionized spectrum is generally observed. The estimated temperature for the central stars of NGC 7027, IC 5117 and IC 4997 is 219000 K \citep{zhang2005}, 120000 K \citep{hyung2001} and 47000 - 59000 K \citep{feibelman1979}, respectively. One can easily see that high-ionization lines such as \ion{He}{ii} or [\ion{Ne}{v}] detected in spectra of planetary nebulae listed above, are not observed in Godzilla.

\item \textbf{Symbiotic stars:}
Symbiotic stars refer to binary systems consisting of a cool giant star transferring mass to an accompanied hot star, mostly white dwarfs. \lyb\ photons from the hot star can excite neutral oxygen atoms in the surrounding gas and induce Bowen fluorescence. \lyb-pumped \ion{O}{i} \lam8449 emission has been observed in many symbiotic starts such as RR Telescopii (RR Tel) \citep{thackeray1955, damineli2001}, AG Pegasi (AG Peg) \citep{ciatti1974, tomov2016}, V1016 Cygni (V1016 Cyg) \citep{strafella1981}, HM Sagittae (HM Sge) \citep{ciatti1977}, and BX Monocerotis (BX Mon) \citep{anupama2012}. \cite{shore2010} also thoroughly examined the \ion{O}{i} \lam1302 and \ion{O}{i}] \lam1641 observed in EG And, Z And, V1016 Cyg, and RR Tel, along with nova RS Oph 1985 in outburst, and concluded that the line strength variation is related to the light curve and outburst activity. Spectra of symbiotic stars listed above all show high-excitation lines including \ion{He}{ii} originating from hot white dwarfs, which are not observed in Godzilla.

\item \textbf{Nova ejecta:}
During a nova eruption, a white dwarf accretes material from its companion star until a thermonuclear runaway occurs. The ejected material is intensely ionized, and as it cools, it produces various emission lines including \ion{O}{i} \lam8449. \lyb-pumped \ion{O}{i} \lam8449 has been observed in several novae in the outburst phase, such as nova Cygni 1975 \citep{strittmatter1977} and nova V4643 Sgr \citep{ashok2006}. In their outburst, novae typically exhibit high-excitation lines such as \ion{He}{ii} and \ion{O}{vi} which are not observed in Godzilla and show rapid spectral evolution on a timescale of days to months.

\item \textbf{Herbig Ae/Be (HAeBe) Stars:} 
These young, pre-main sequence stars, defined by \cite{herbig1960}, are surrounded by accretion disks. UV radiation from these stars, as well as shocks within the accretion disks, can induce the Bowen fluorescence. \cite{mathew2018} argued that \lyb-pumping is the primary mechanism responsible for \ion{O}{i} \lam8449 line observed in HAeBe stars. Although HAeBe stars show many spectral similarities to Godzilla, they have less in common compared to $\eta$ Car and the Weigelt blobs, which will be discussed next. Figure 2 of \cite{mathew2018} shows that in HAeBe stars, the \ion{O}{I} $\lambda8449/\lambda7776$ ratio can reach up to about $\sim25$ – 50 in extreme cases, whereas in Godzilla, this ratio is $\sim110$. While \ion{O}{I} \lam7776 is fairly prominent in many HAeBe stars, this line is not detected in Godzilla. Furthermore, due to the presence of a stellar disk, H$\alpha$ in HAeBe stars often shows a double-peak profile, and even when a single peak is present, it lacks the broad wings observed in Godzilla \citep[Fig. 3]{carmona2010}. 

\end{itemize}

\subsubsection{LBV - $\eta$ Car and the Weigelt blobs} \label{sssec:lbv_scenario}
We find that the peculiar \ion{O}{i} \lam8449 source in Godzilla is best explained as an analog of the Luminous Blue Variable (LBV) star $\eta$ Car and the Weigelt blobs, given its spectral characteristics. In Section~\ref{ssec:bowen}, we have shown that the strong, narrow permitted oxygen line \ion{O}{i} \lam8449 has most likely arisen from a \lyb\ pumping mechanism. In Section~\ref{sssec:excluded_candidates}, we have reviewed a number of known astronomical sources of \ion{O}{i} \lam8449 emission, and have shown that the combination of a strong \ion{O}{i} \lam8449 line and the absence of the \ion{O}{i} \lam7776 line is a rare feature, observed only in Weigelt blobs among objects that do not exhibit high-ionization lines. Moreover, as mentioned in Section~\ref{ssec:temden}, we know that dense gas of $n_{\rm e} \gtrsim 10^6 \:\rm cm^{-3}$ exists in close vicinity of Godzilla \citep{vanzella2020}, although our optical density diagnostics only offer limited constraints on low density gas. Existence of high density gas and the strong fluorescent line emission is evocative of the Weigelt blobs in the $\eta$ Car systems explained in Section~\ref{sec:introduction}. 

The Weigelt blobs are dense gas condensations that are primarily neutral, with an ionized surface layer facing the stars \citep[See Figure 1. in][for a sketched out overview]{johansson2005}. The \ion{O}{i} in the neutral condensations is shielded by hydrogen from ionisation, but exposed to \lyb\ photons emitted from the \ion{H}{ii} surface of the blob, as well as the stellar wind, populating the 3d$^3$D level of the neutral oxygen atoms \citep[Figure 4]{johansson2005}. This level has an excitation energy very close to the resonant energy of \lyb\ photons and can be pumped by it, if \lyb\ is sufficiently strong - a mechanism known as PAR. \citet[Figure~1]{shore2010} show a different view of this same cascade. Once the 3d$^3$D level is populated, it decays to 3p$^3$P level through a $\lambda = 11286$ \AA\ transition which is outside the wavelength range of these observations. Subsequently, they continue to cascade down from 3p$^3$P to 3s$^3$S$_1$ through the $\lambda = 8449$ \AA\ transition, and from there into \ion{O}{i}]\lam1641 and the $\lambda = $ 1302, 1304, 1306 triplet. We note that all of these lines are strongly detected in the MagE and NIRSpec observations of Godzilla. In the Weigelt blobs, as oxygen in the interior of the blob remains neutral, it results in a special situation where we observe strong \lyb-pumped \ion{O}{i} \lam8449, but not the recombination line \ion{O}{i} \lam7776. The clear detection of the full \lyb-pumped cascade, combined with the non-detection of \ion{O}{i} \lam7776 (Section~\ref{ssec:bowen}, \autoref{fig:OI_MagE}) is the smoking gun that permitted \ion{O}{i} lines in Godzilla has arisen from dense, Weigelt blob-like gas condensations.

In the Weigelt blobs, [\ion{Ne}{iii}] is only observed at times when the blob is exposed to the secondary star, which is hotter than the LBV star. We can explain the significant dust attenuation, the strong Bowen fluorescent emission, and the [\ion{Ne}{iii}] emission line simultaneously, if Godzilla is a binary system including a massive evolved LBV-like star and a hotter star, analogous to $\eta$ Car. It is a plausible scenario considering that 70\% of massive stars are affected by binary interaction, and 50\% have companions \citep{weis2020}. It is also noteworthy that \cite{smith2015} and \cite{smith2019} suggested that LBVs result from close binary evolution. Alternatively, the properties listed above can be explained without invoking a binary scenario, if the gas condensations in Godzilla are exposed to more intense radiation of an LBV star compared to the Weigelt blobs. Assuming the LBV star has a similar temperature to $\eta$ Car, this would be possible if the blobs were located more closely to the star than the Weigelt blobs to $\eta$ Car. These two scenarios could potentially be distinguished by observing periodic variability in line emission. However, depending on the inclination and viewing angle, spectral variability might not be detected, even if it is indeed a binary system.

We would also like to discuss the magnification and brightness issues raised in previous studies. \cite{diego2022} has noted that with $\mu \approx 7000$, the true brightness of Godzilla would match that of $\eta$ Car during its Great Eruption. Despite the high resolution of JWST/NIRSpec, we do not detect any signs of an eruption such as a very broad H$\alpha$ ($\sim 10^4$ \kms) or P Cygni profiles. If we adopt the values suggested by \cite{diego2022}, this implies that the magnification factor for Godzilla must exceed 7000 if the source has the same magnitude as $\eta$ Car in its non-eruption phase. However, it should be noted that the analysis by \cite{diego2022} is based on the assumption that Godzilla is a single stellar source. If Godzilla is indeed composed of multiple stars, the constraints on the magnification factor could be relaxed.  

The overall picture we propose is as follows: Godzilla is part of clump 4, comprising 10 - 25\% of its stellar light, depending on the lens models and the images in comparison. It could be a small group of just a few stars, or tens or even hundreds of stars. These stars contribute significant or even dominating fraction of the stellar continuum. Within this group of stars, a peculiar \ion{O}{i} \lam8449 source is present, which is best explained as a source similar to the Weigelt blobs in the $\eta$ Car system. The spectra of Godzilla resemble those of $\eta$ Car in its quiescent phase, with nearby dense gas condensations exposed to a hotter source, either due to a presumable hotter companion or their potentially closer location to the star. We believe that the emission lines observed in Godzilla are primarily from this dense gas condensation, though some lines, such as redshifted [\ion{O}{ii}] lines, may originate from other regions (\autoref{fig:center_temden}). Our scenario is a kind of hybrid model, differing from both \cite{diego2022}'s interpretation of Godzilla as a single object and \cite{pascale2024}'s interpretation as a star cluster, in the sense that the stellar continuum is affected by multiple stars, while a single object analogous to $\eta$ Car dominates the emission line. 
In the next section, we further examine the similarities and differences between \ion{O}{i} \lam8449 in Godzilla and the Weigelt blobs.




\subsection{Comparison to Bowen fluorescent lines observed in $\eta$ Carinae} \label{ssec:bowen_discussion}
\subsubsection{A possible astronomical laser effect} \label{sssec:laser}
\cite{johansson2005} discuss the possibility of astrophysical laser effects of \ion{O}{i} \lam8449 in the Weigelt blobs. 
As the decay of the lower 3s$^3$S$_1$ level to the ground level is faster than the decay of the upper 3p$^3$P level, population inversion occurs. \ion{O}{i} \lam8449 from Godzilla is due to \lyb\ pumping, so we expect it to show an inverted population, but it does not necessarily mean that it is a laser since we cannot guarantee stimulated emission without information on the size of the blob and amplification of the \ion{O}{i} \lam8449 line. For observational confirmation, \cite{johansson2005} suggests looking for  sub-Doppler width, i.e. the line width narrower than the width caused by Doppler broadening, of the 8449 line. 

\lya-pumped iron lines can provide us an additional clue about the sub-Doppler width of the \ion{O}{i} \lam8449 line. Along with the \lyb-pumped \ion{O}{i} \lam8449, we detect \lya-pumped \ion{Fe}{ii} \lam8490 with S/N of 8 (see \autoref{tab:red_flux}). 
Interestingly, another \lya-pumped iron line, \ion{Fe}{ii} \lam8453, which arise from the same mechanism thus should appear is not detected in the fits using [\ion{O}{iii}]\lam\lam4960,5008 as kinematic template (\autoref{tab:red_flux}). This line is blended with \ion{O}{i} \lam8449, and we suspect that its flux may have been included in the \ion{O}{i} \lam8449 flux since the \ion{O}{i} \lam8449 line is narrower than other lines. We fitted the fluorescent lines again, using only one component and  relaxing the width constraint. The result is shown in \autoref{tab:red_flux} with asterisk. We measured FWHM of 106$\pm$5 \kms\ for the fluorescent lines and it increased S/N of \ion{Fe}{ii} \lam8453 from $\sim 1.8$ to $\sim 12$. However, along with the lack of information on the thermal broadening in this object, the spectral resolution of NIRSpec does not allow sufficiently strong constraints on the line profile to confirm or falsify if the \ion{O}{i} \lam8449 emission observed in Godzilla is an astronomical laser. 

\subsubsection{Flux discrepancy in \lya-pumped lines} \label{sssec:lya_pumped}
The relative strengths of \lya-pumped \ion{Fe}{ii} lines compared to the \lyb-pumped \ion{O}{i} \lam8449 line is significantly smaller than observed in the Weigelt blobs \citep[see][Figure 5.4]{davidson2012}. In the Weigelt blobs, \ion{Fe}{ii} fluorescent lines are much stronger than \ion{O}{i} \lam8449. This might imply an abundance difference, which is not surprising considering that Godzilla is at $z=2.37$ and the Weigelt blobs are in our Galaxy, where the metal enrichment has been enhanced by Type Ia SN. Several observations have already suggested significant alpha enhancement, i.e. iron deficiency, in $z\sim 2-3$ galaxies \citep{steidel2016c,topping2020a,topping2020,cullen2021}. Alternatively, it might be a circumstantial evidence of a \ion{O}{i} laser and significant amplification in that line.

\section{Conclusions} \label{sec:conclusion}
We report the results from Cycle~1 JWST/NIRCam and NIRSpec observations of the object nick-named ``Godzilla'' in the Sunburst Arc. The Sunburst Arc is the brightest known gravitationally lensed galaxy at $z=2.37$ and consists of 12 full or partial images of the source galaxy. However, Godzilla only appears in one of these, despite being one of the brightest elements of that image. While the absence of counterimages of Godzilla has been explained with extreme magnification \citep{diego2022, sharon2022}, there is a still ongoing debate on the nature of the object \citep{vanzella2020, pascale2024}. It has been suggested that Godzilla might be a rare stellar object called a Luminous Blue Variable (LBV) based on the detection of uncommon Bowen fluorescent lines \citep{vanzella2020} and the luminosity constraints from the lens model \citep{diego2022}. However, ground-based telescope observations from previous studies do not have a spatial resolution to test this at such a high redshift. In this work, we have extracted integrated spectra of Godzilla from NIRSpec IFU spaxels containing the central bright object (Center) and four regions (NE, SE, NW, SW) surrounding it (\autoref{fig:five_regions}), and measured the flux of emission lines using one to three component Gaussian profiles depending on the regions and lines.
Our main findings are as follows:
\begin{enumerate}
    \item We detected 57 emission lines from JWST/NIRSpec spectrum of the bright central region (Center), tabulated in \autoref{tab:blue_flux} and \autoref{tab:red_flux}. These include auroral lines of four species ([\ion{N}{ii}]\lam5756, [\ion{O}{ii}]\lam\lam7322,7332, [\ion{O}{iii}]\lam4363, and [\ion{S}{iii}]\lam6314), and strong permitted \ion{O}{i} \lam8449. 
    \item NIRCam observations extend the baseline of observations of Godzilla to $\sim 1.5$ years in the rest-frame; the source has an approximately flat light-curve over that time, and no counterimages have been detected, strengthening the case against the Type IIn supernova scenario or any similarly short-lived transient event (\autoref{fig:nircam_image}). Emission line profile comparison between Godzilla, the P knots (a pair of small clumps next to Godzilla) and the LCE (\autoref{fig:p_knots_lines}) supports the microlensing scenario suggested by \cite{diego2022}. We suggest a newly detected low surface brightness galaxy near Godzilla (\autoref{fig:perturber}) as a possible perturber creating the critical curve crossing on top of Godzilla. We agree that Godzilla only appears at its current position since it is extremely magnified  (Section~\ref{ssec:lensing}).
    \item Based on the similarity in color (\autoref{fig:color_color}) and the strong \ion{O}{i} \lam8449 emission (\autoref{fig:O1_O3_map}), we concluded that the images of clump 4 are containing Godzilla. We compared the flux of the stellar continuum and the \ion{O}{i} \lam8449 emission line between Godzilla and the clump 4 images (\autoref{tab:conterimages_magnification}). Compared to images 4.4 and 4.9, while Godzilla is 35 times and 94 times brighter in \ion{O}{i} \lam8449, it is only 8 and 10 times brighter in the continuum, respectively, implying that Godzilla contains 10–25\% of the the total stellar light of the clump 4 (\autoref{fig:conceptual_figure}). Depending on the lens models used, the magnification factor of Godzilla can vary widely, ranging from $\approx600$ to $\approx25000$.
    \item We showed that \ion{O}{i} \lam8449 is most likely produced through \lyb-pumping (Section~\ref{ssec:bowen}), reporting the UV counterpart of the \lyb-pumped \ion{O}{i} triplet detected in a Magellan/MagE spectrum, along with \ion{O}{i}]\lam1641 (\autoref{fig:OI_MagE}). By examining possible scenarios (Section~\ref{sssec:excluded_candidates}), we concluded that the presence of this \lyb-pumped cascade implies the existence of dense gas condensations in close vicinity of the energy source, which is analogous to the Weigelt blobs near Eta Carinae ($\eta$ Car), an LBV star in Milky Way (Section~\ref{sssec:lbv_scenario}). This conclusion is in line with the previous detection of \lya-pumped Bowen lines \citep{vanzella2020}.
    We also discussed the hypothesized natural laser effect in \ion{O}{i} \lam8449 previously suggested in the literature \citep{johansson2005}; but while we find the suggestion plausible, we have not found basis in the data to confirm or refute a contribution to the line emission from such a natural laser (Section~\ref{ssec:bowen_discussion}).
    
    \item Godzilla and the surrounding regions are very dusty (\autoref{tab:dust_factors}) and the broadest component shows a moderately high velocity dispersion of 435 - 547 \kms\ (\autoref{tab:kinematics}). Based on spatially varying line profiles (\autoref{fig:five_regions}), dust reddening, temperature and density (\autoref{fig:center_temden}, \autoref{fig:temden_wings}), and velocity, we postulate that this object is surrounded by multi-phase circumstellar or interstellar medium. 
    \item We find that an LBV with a hotter companion (analogous to $\eta$ Car) best qualitatively explains our observations simultaneously while requiring the least extreme magnification. Gas condensations located more closely to the source, thus exposed to the harder radiation compared to the Weigelt blobs, can be an alternative scenario. Considering (1) that the measured maximum velocity of the broadest component of H$\alpha$ in Godzilla is an order of magnitude smaller than that of $\eta$ Car during the Great Eruption (\autoref{fig:Ha_very_broad}, Section~\ref{ssec:kinematics}) and (2) the non-detection of P Cygni profiles, we conclude that the source is not going through an eruption phase. A spectroscopic monitoring program of Godzilla might help determine whether there is periodic variations in nebular emission lines, such as the 5.54-year spectroscopic period seen for $\eta$ Car, although the variation effect could be highly affected by the inclination and viewing angle.
\end{enumerate}

``Godzilla'' is a truly remarkable object.
In addition to existing clues in the literature, we find strong spectroscopic evidence pointing to the object containing an $\eta$ Car analog, with very dense, neutral gas condensations exposed to more intense emission, possibly with a smaller, hot, blue companion. 
Future investigation of possible spectral variability and refined lens models may help further constrain the nature of Godzilla.

\begin{acknowledgements}
The authors would like to thank Seán Brennan, Ragnhild Lunnan and Yang Hu, Stockholm University, for helpful discussions about late stage evolution of massive stars. This work is in part based on observations made with the NASA/ESA/CSA James Webb Space Telescope. The data were obtained from the Mikulski Archive for Space Telescopes at the Space Telescope Science Institute, which is operated by the Association of Universities for Research in Astronomy, Inc., under NASA contract NAS 5-03127 for JWST. These observations are associated with program \#2555.
Support for program \#2555 was provided by NASA through a grant from the Space Telescope Science Institute, which is operated by the Association of Universities for Research in Astronomy, Inc., under NASA contract NAS 5-03127.
ER-T is supported by the Swedish Research Council grant Nr. 2022-04805\_VR. This work makes use of the software packages Astropy \citep{theastropycollaboration2013,theastropycollaboration2018,theastropycollaboration2022}, Matplotlib \citep{hunter2007}, NumPy \citep{harris2020}, Uncertainties Python package by Eric O. LEBIGOT, Jupyter notebook \citep{kluyver2016}, LMFIT \citep{newville2014} and PyNeb \citep{luridiana2015a}. 
\end{acknowledgements}

%
   \bibliographystyle{aa} 
   \bibliography{bibliography} 
%

\begin{appendix} 
\section{Spectral line identification}
   \begin{figure}[!htp]
    \onecolumn \includegraphics[height=0.9\textheight]{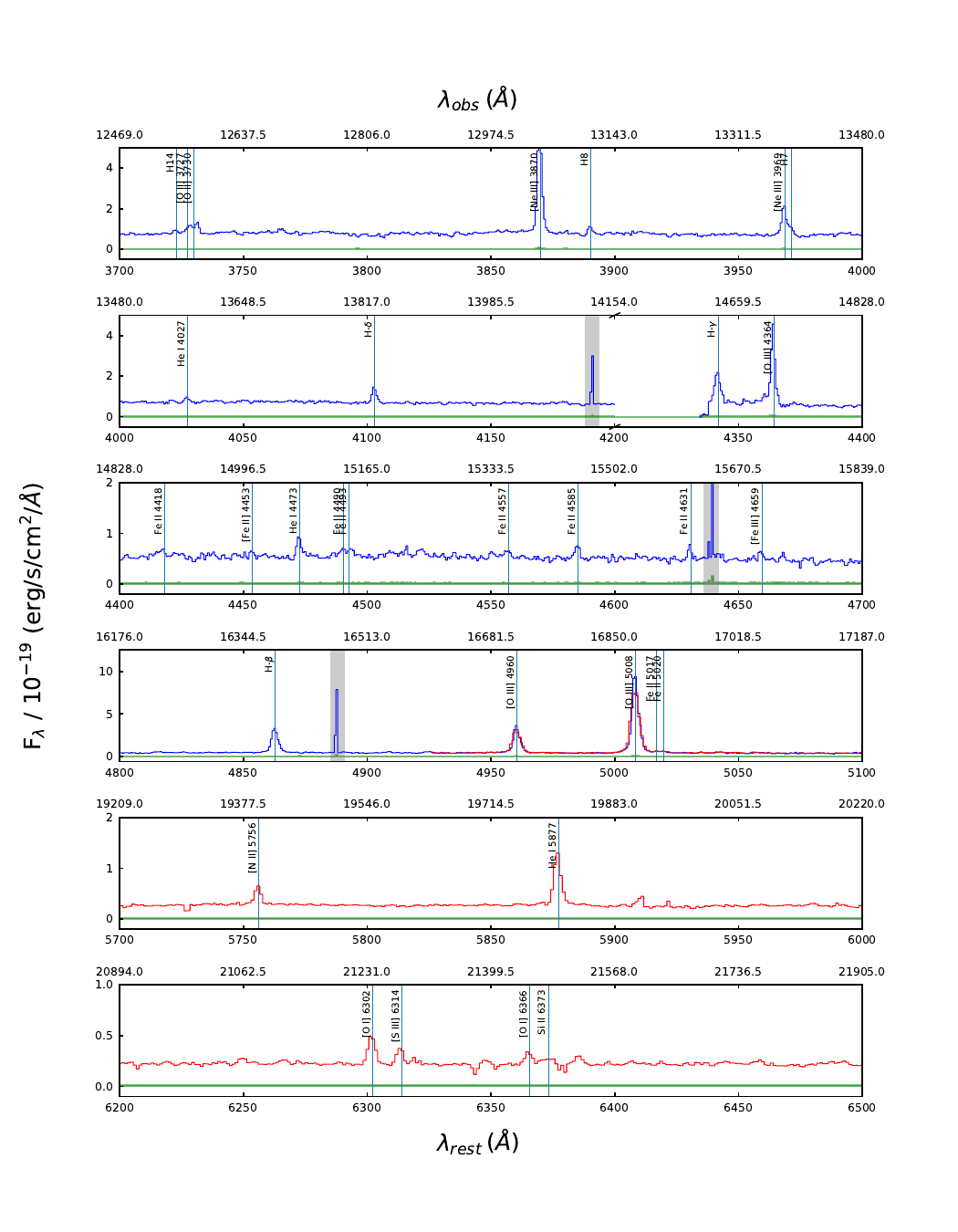}
   \caption{One-dimensional spectrum from the bright central region of Godzilla. Observed-frame flux densities from the G140H/F100LP and G235H/F170LP gratings are shown in blue and red solid lines, respectively. Standard deviation is shown with green area and hot/noisy pixels are marked with grey shades.}
              \label{fig:spectra_1}%
    \end{figure}
    \twocolumn

   \begin{figure*}[!htp]
   \centering
   \onecolumn \includegraphics[height=0.9\textheight]{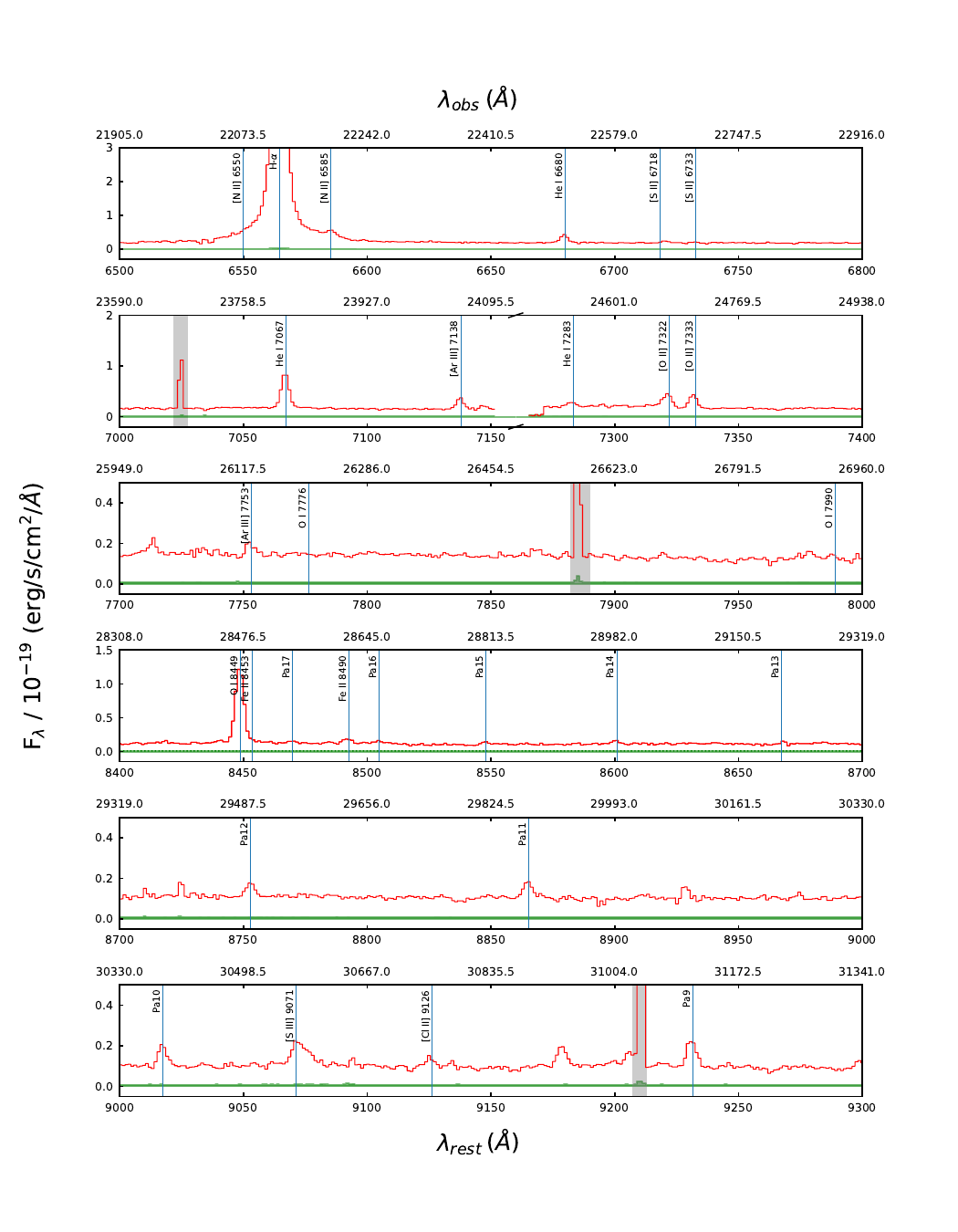}
   \caption{Same as Figure~\ref{fig:spectra_1}, continuing to 9300 \AA.}
              \label{fig:spectra_2}%
    \end{figure*}
    \twocolumn

\section{Properties in surroundings}
   \begin{figure*}[!htp]
   \centering
   \onecolumn \includegraphics[width=\textwidth]{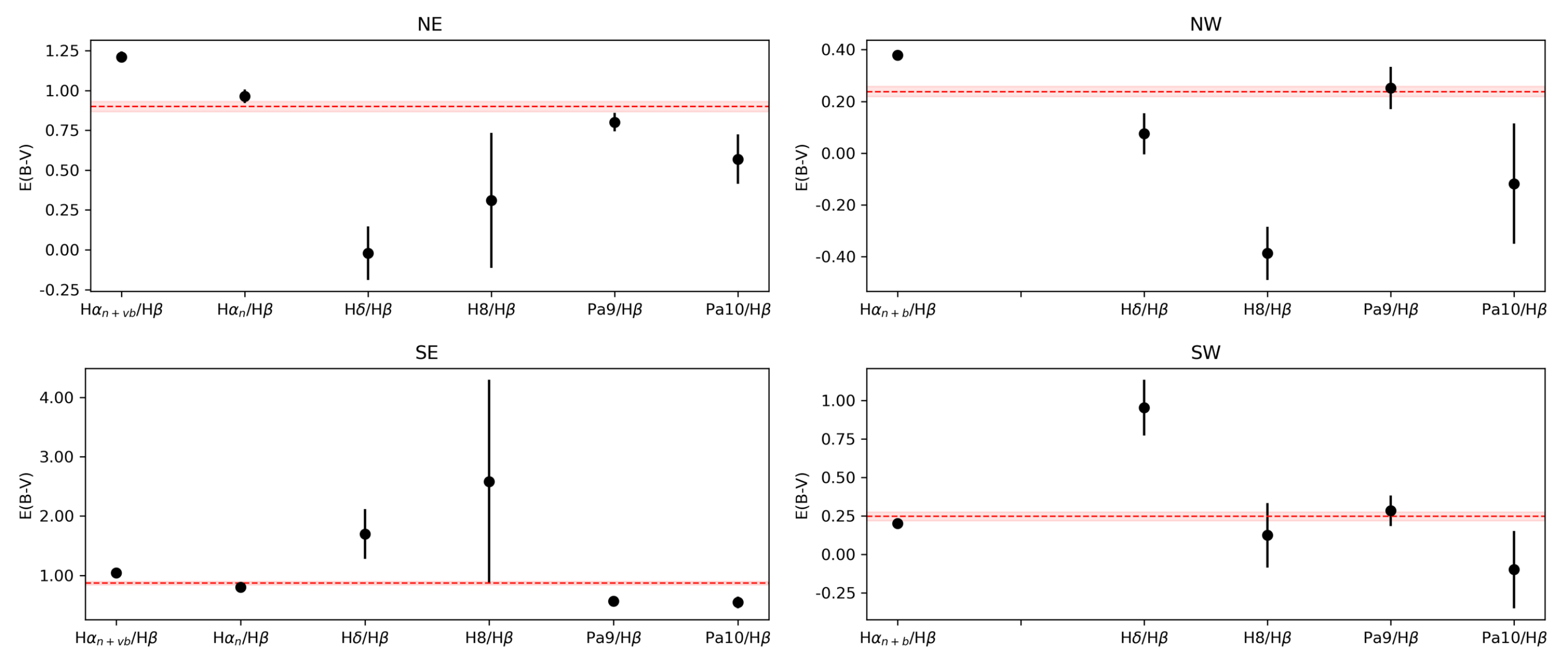}
   \caption{$E(B-V)$ computed for the NE, SE, NW, SW regions in the same manner as \autoref{fig:ebv_Center}. The red-dashed horizontal lines show the inverse error weighted averages.}
              \label{fig:ebv_wings}%
    \end{figure*}
    \twocolumn

   \begin{figure*}[!htp]
   \centering
   \onecolumn \includegraphics[width=\textwidth]{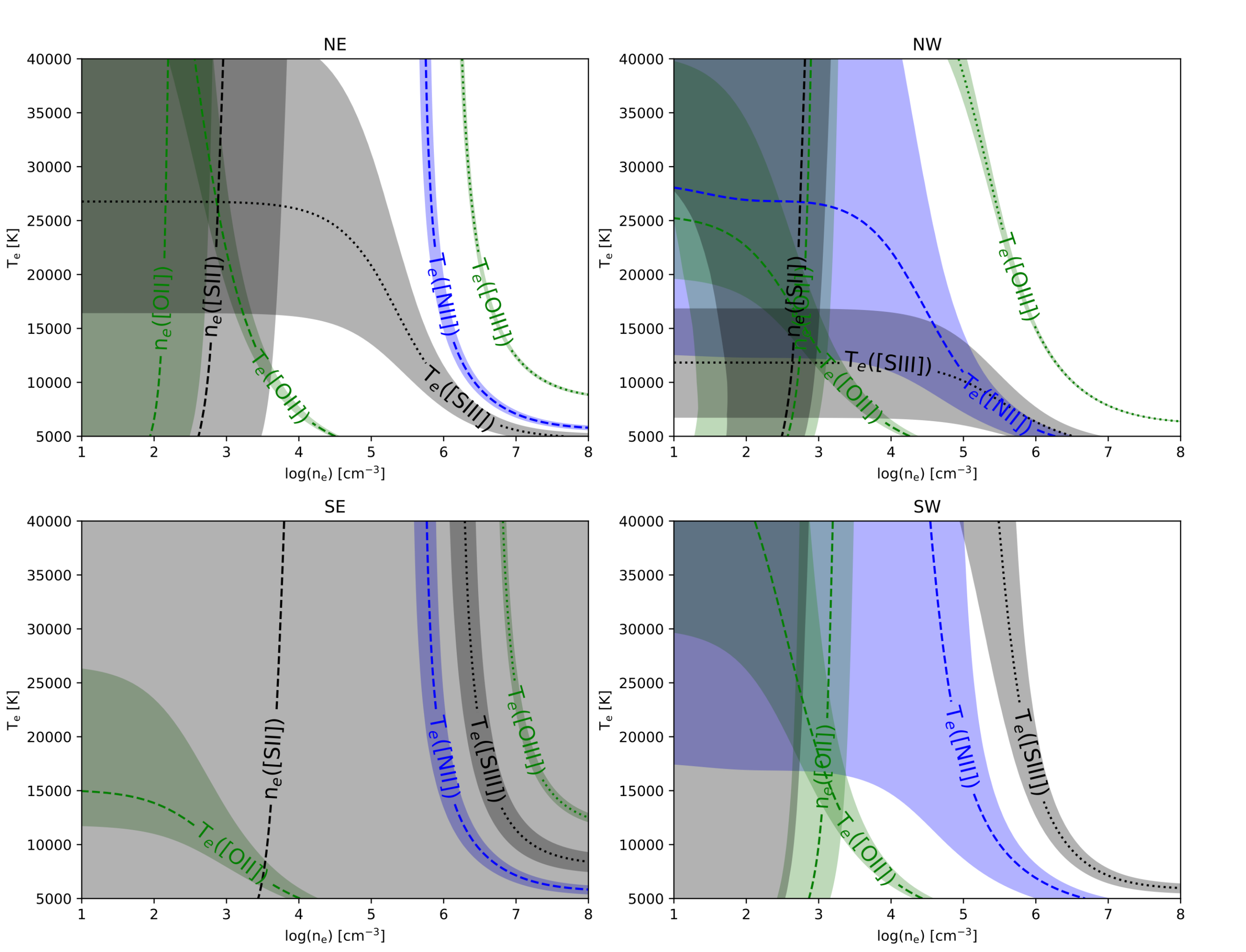}
   \caption{Joint temperature and density solutions computed with PyNeb for the NE, SE, NW, SW regions, using the same nebular lines as found in \autoref{fig:center_temden}.}
              \label{fig:temden_wings}%
    \end{figure*}
    \twocolumn

\section{\ion{O}{i} \lam8449 map normalized by H$\beta$}
\begin{figure*}[!htp]
   \centering
\onecolumn \includegraphics[width=\textwidth]{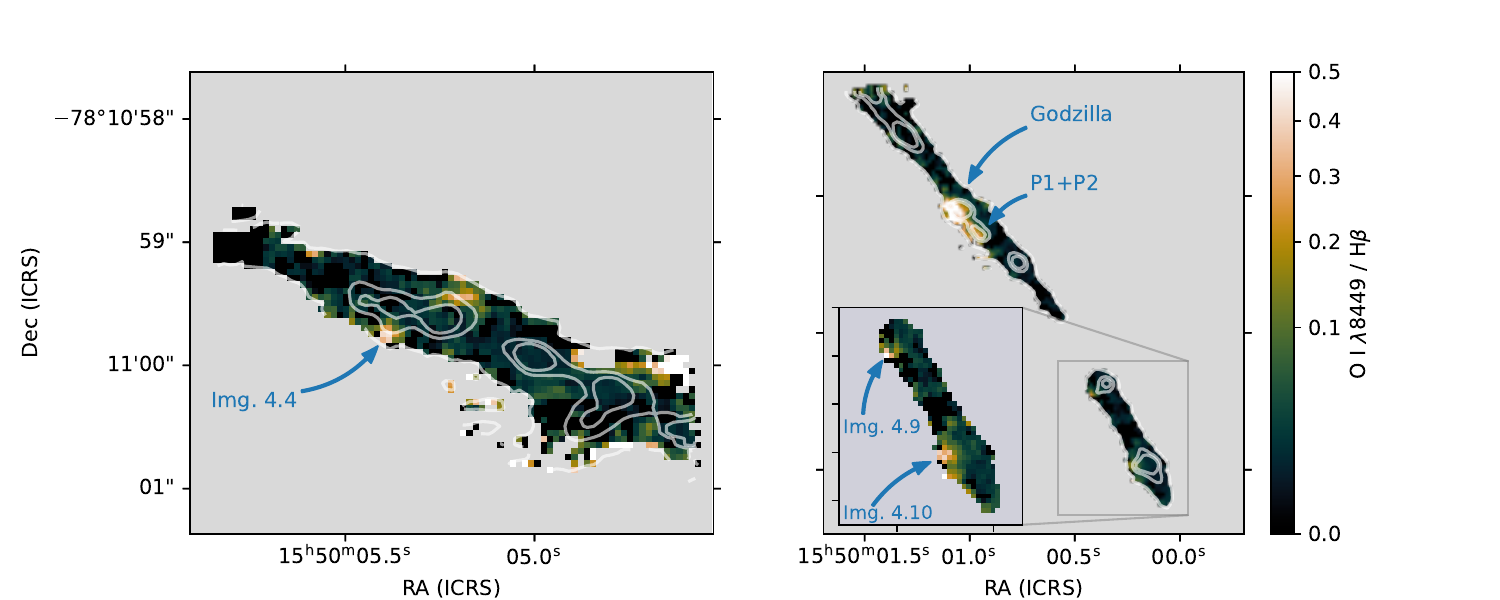}
   \caption{\ion{O}{i} \lam8449/H$\beta$ map for Pointing 1 (left) and Pointings 2+3 (right), with the stellar continuum overlaid as contours. The excess in \ion{O}{i}/H$\beta$ is observed in Godzilla, the P knots and the counterimages of the clump 4 as in the [O III] normalization map (\autoref{fig:O1_O3_map}).}\label{fig:O1_Hb_map}
    \end{figure*}
    \twocolumn

\end{appendix}

\end{document}